\documentclass[%
 aip,
 amsmath,amssymb,
 reprint,%
]{revtex4-1}

\usepackage{graphicx}
\usepackage{dcolumn}
\usepackage{bm}
\usepackage{float}
\usepackage[table,xcdraw]{xcolor} 
\usepackage[utf8]{inputenc}
\usepackage[T1]{fontenc}
\usepackage{mathptmx}
\usepackage{etoolbox}
\usepackage{nomencl}
\usepackage{multirow}
\usepackage{tabularx}
\usepackage{hyperref}
\hypersetup{
    colorlinks=false, 
    linktoc=all,     
    linkcolor=blue,
    colorlinks=false,
}
\hypersetup{hidelinks}
\makenomenclature

\makeatletter
\def\@email#1#2{%
 \endgroup
 \patchcmd{\titleblock@produce}
  {\frontmatter@RRAPformat}
  {\frontmatter@RRAPformat{\produce@RRAP{*#1\href{mailto:#2}{#2}}}\frontmatter@RRAPformat}
  {}{}
}%
\makeatother
\begin{document}

\preprint{AIP/123-QED}

\title[Accelerated GEP to Predict Wall Pressure Spectra beneath Turbulent Boundary Layers.]{Artificial Neural Networks and Guided Gene Expression Programming to Predict Wall Pressure Spectra Beneath Turbulent Boundary Layers}

\author{Nachiketa Narayan Kurhade}
\author{Nagabhushana Rao Vadlamani$^*$}%
 \email{nrv@iitm.ac.in}
\affiliation{Department of Aerospace Engineering\char`,{} Indian Institute of Technology Madras\char`,{} 600036\char`,{} India}

\author{Akash Haridas}
\affiliation{Department of Computer Science\char`,{} University of Toronto\char`,{} Toronto ON M5S3G4
}

\date{\today}

\begin{abstract}
This study evaluates the efficacy of two machine learning (ML) techniques, namely artificial neural networks (ANN) and gene expression programming (GEP) that use data-driven modeling to predict wall pressure spectra (WPS) underneath turbulent boundary layers. Different datasets of WPS from experiments and high-fidelity numerical simulations covering a wide range of pressure gradients and Reynolds numbers are considered. For both ML methods, an optimal hyperparameter environment is identified that yields accurate predictions. ANN is observed to be faster and more accurate than GEP with an order of magnitude lower training time and logarithmic mean squared error ($lMSE$), despite a higher memory consumption. Novel training schemes are devised to address the shortcomings of GEP. These include (a) ANN-assisted GEP to reduce the noise in the training data, (b) exploiting the low and high-frequency trends to guide the GEP search, and (c) a stepped training strategy where the chromosomes are first trained on the canonical datasets followed by the datasets with complex features. When compared to the baseline scheme, these training strategies accelerated convergence and resulted in models with superior accuracy ($\approx 30\%$ reduction in the median $lMSE$) and higher reliability ($\approx 75\%$ reduction in the spread of $lMSE$ in the interquartile range). The final GEP models captured the complex trends of WPS across varying flow conditions and pressure gradients, surpassing the accuracy of Goody's model.

\end{abstract}

\maketitle

\section{\label{sec1}Introduction}
Turbulent boundary layers (TBLs) developing over the surfaces induce wall-pressure fluctuations. These dynamic loads are highly undesirable as they increase structural vibrations and noise. Fatigue failure due to aeroacoustic loads is a critical problem in wind turbine and gas turbine blades \cite{Roger2005,Bull1996,Hambric2004,Avallone2018,Tang2019}. Aerodynamic noise due to attached or separated TBLs contributes significantly towards medium and high-frequency sound pressure levels (SPL) of the cabin \cite{Wilby1972,Wilby1989}. Acoustic loads due to wall-pressure fluctuations are much more severe over the skin panels of high-speed vehicles. The amplitude and frequency of these loads can be significantly higher in certain regions which can potentially damage the vehicle. 

Estimating the wall-pressure fluctuations beneath TBLs is hence crucial to facilitate the structural and aerodynamic design process. These fluctuations can either be estimated from high-fidelity eddy-resolving simulations or can be directly measured from experiments. Both these approaches are prohibitively expensive. Although eddy-resolving methods like Direct Numerical Simulations (DNS)\cite{Choi1990,Deuse2019,Wu2018} and Large Eddy Simulations (LES)\cite{Cohen2018,Christophe2009,saiteja} can accurately quantify the spatiotemporal variation of acoustic loads, they are computationally expensive to cater to the highly iterative design process. On the other hand, most of the experiments in the literature are confined to measuring wall pressure spectra (WPS) of TBLs developing over simple geometries like a flat plate subjected to zero, favorable, or adverse pressure gradients \cite{Salze2014,goody2000surface,blake1970turbulent,Farabee1991,van2018} (ZPG, FPG, and APG). 

One of the widely used approaches to overcome these limitations is to couple the empirical models of WPS with the mean boundary layer characteristics obtained from the Reynolds Averaged Navier-Stokes (RANS) simulations to predict acoustic loads \cite{Rozenberg2012}. These models rely on scaling laws based on the inner and outer regions of the boundary layer to estimate the sound pressure levels. Several empirical models for WPS were developed based on the theoretical foundations of Lighthill\cite{Lighthill1952,Lighthill1954} and Kraichnan\cite{Kraichnan1956}. Blake\cite{blake} integrates the Poisson equation for pressure fluctuation beneath a TBL. As noted by Grasso \emph{et al.} \cite{Grasso2019}, the theoretical model involves repeated integration in multiple dimensions, and the computational cost increases exponentially with the number of dimensions. An alternate approach involves correlating the non-dimensionalized SPL distribution with boundary layer parameters like thickness, wall-shear stress, freestream velocity, etc. This approach, as reflected in the works of Chase\cite{Chase1980} and Howe\cite{howe} has an evolutionary advantage as the subsequent models continued to modify the same baseline empirical structure to account for newer observations. Goody\cite{Goody2004} further modified the Chase-Howe model to include Reynolds number effects and capture the frequency dependence. Goody's model is designed to predict the WPS of TBLs under ZPG with a \(\omega^2\) growth at low frequencies (in line with Kraichnan-Philips\cite{Bull1996} theorem),  \(\omega^{-1}\) decay in the inertial range (in line with Bradshaw prediction\cite{Bradshaw1967}) and \(\omega^{-5}\) decay at the high-frequencies (in line with Blake's observations \cite{blake}). Subsequent works of Kamruzzaman \textit{et al.}\cite{Kamruzzaman2015}, Rozenberg \textit{et al.}\cite{Rozenberg2012}, Hu\cite{Hu2018} and Lee \cite{Lee2018} extended Goody’s model to account for pressure gradients. Ritos \textit{et al.}\cite{ritos2019wall} recognised that the existing models fall short in predicting the WPS beneath supersonic and hypersonic flat plate TBLs under ZPG conditions and modified Goody's model to account for the compressibility effects. Wind tunnel experiments of Thomson and Rocha\cite{thomson2022semi} measured wall pressure fluctuations of TBL on a flat plate under FPG. They further tuned the parameters of the universal spectrum model (which is inspired by Goody's model) to incorporate FPG effects and improve predictions at high-frequencies. Thomson and Rocha\cite{thomson2021comparison} compare the performance of Goody's model on both the wind tunnel data and flight test data. They highlight that the model accurately predicts the WPS of the former while underpredicts the latter and accordingly updated Goody's model to accurately fit the flight test data. Accounting for the different flow conditions and pressure gradients, Fritsch \textit{et al.}\cite{fritsch_2023} developed a semi-empirical WPS model using numerical optimization algorithms. The predictions are shown to be sensitive to the values of constants involved and the optimized set is bound to evolve with an ever-evolving dataset. All these empirical models, however, were developed based on the respective datasets and hence their accuracy suffers when employed for `extrapolated' flow conditions.


With increasing computational power and the databases of wall pressure spectra data published in public domains, robust wall pressure spectra models can be developed by exploiting the entire database. Additionally one can identify the range of TBL parameters covered by these databases \cite{Dominique2022}. These insights can then be used by modern machine learning (ML) algorithms to solve this regression problem. Haridas and Vadlamani\cite{Haridas2021} and Dominique \textit{et al.}\cite{Dominique2022} used artificial neural networks (ANN) for this purpose and obtained better predictions than any of the empirical models discussed above. The study conducted by Haridas and Vadlamani \cite{Haridas2021} used subsonic and supersonic datasets to train their ANN model. This approach also enabled Dominique \emph{et al.} to quantify the confidence in their predictions and identify the input space where more data will be helpful in improving the predictions. ANN being the universal function approximator \cite{Hornik1989} is indeed best suitable for the job as it can be retrained with little effort on the new datasets. The downside with ANN is twofold: (a) It results in a highly non-linear function that is difficult to translate to a conventional functional form and (b) It provides little insight into the underlying physics of the problem \cite{Haridas2021}. On the other hand, Gene expression programming (GEP), a symbolic machine learning algorithm, operates on the input variables through generations to evolve into an analytical expression that tries to approximate provided data. Dominique \emph{et al.}\cite{Dominique2021} showed the capabilities of this technique to predict the WPS models that are close to the established empirical correlations in addition to discovering the new dependencies that were not considered earlier. GEP however suffers from converge issues. Also, for a given set of input features, the algorithm produces different analytical expressions, although their structures and predictions can be very similar.


The objective of the present work is multifold. We first compare the strength and weaknesses of ANN and GEP (with an optimized hyperparameter environment) against the same input data assessing the time taken to converge, and the computational resources utilized. Subsequently, we propose strategies (like ANN-assisted training incorporating physical insights) to accelerate the training of the GEP algorithm and mitigate its convergence issues. Finally, we demonstrate the efficacy of the training strategies and the robustness of the models predicted using the guided GEP approach.

The manuscript is structured as follows: We introduce the regression problem and the relevant WPS models in Section \ref{sec1:subsec1}. Section \ref{sec2} presents a brief description and analysis of the input dataset and a short overview of ANN and GEP. We present the optimum hyperparameter environment and compare the computational efficacy of ANN and GEP in Section \ref{sec3}. Strategies to accelerate GEP training to mitigate the convergence issues are addressed in Section \ref{ch4}. Lastly, Section \ref{conclusions} concludes the key findings of the study.

\section{\label{sec1:subsec1} Characterizing Turbulent boundary layer and wall pressure spectrum}

\begin{figure}
\includegraphics[width=\linewidth]{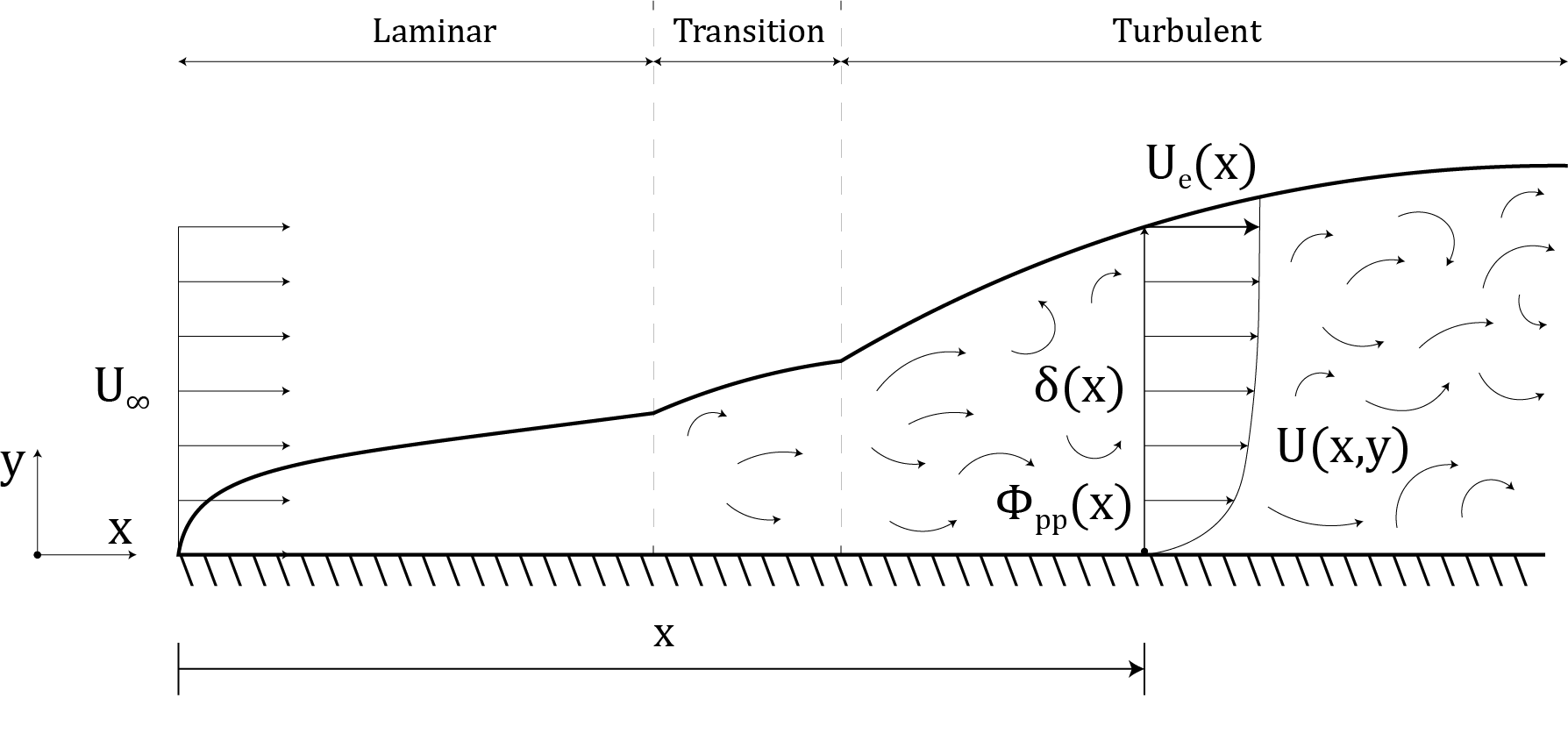}
\caption{\label{fig:boundary_layer}Typical flat plate boundary layer}
\centering
\end{figure}

As mentioned in the introduction, the empirical models of WPS rely on the mean boundary layer characteristics to predict acoustic loads. All the TBLs have a negligible zero pressure gradient in the wall-normal direction ($\partial_{y}p = 0$) but can be subjected to a considerable gradient ($\partial_xp$) in the streamwise direction. Figure ~\ref{fig:boundary_layer} sketches a typical velocity profile $U(y)$ at a distance $x$ downstream of the leading edge for a general case where the equilibrium velocity at the boundary layer edge changes with $x$ as $U_e(x)$. The local boundary layer thickness ($\delta(x)$) is usually defined as the wall-normal distance at which $U = 0.99U_e(x)$. In this study, following Dominique \emph{et al.}, we use an alternate definition for boundary layer thickness based on the pseudo-velocity\cite{Deuse2019} $U^*(x,y) = -\int_{0}^{y} \Omega_z(x,\xi) \,d\xi$ to account for the effect of pressure gradients on the outer flow velocity. Here, $\Omega_z$ is the spanwise component of the mean vorticity $\Omega = \nabla \times \overline{U}$, $\overline{U}$ being the mean velocity vector and $\xi$ is the length in wall-normal direction.

At the same point $x$ downstream of the leading edge, a probe can be placed on the wall to record the pressure fluctuations due to TBL. The unsteady pressure signals $p(\tau)$ can be decomposed into its mean and fluctuating components as follows:

\begin{equation}\label{-105}
    p(x,\tau) = \overline{p}(x,\tau) + p^\prime(x,\tau)
\end{equation}

The point-wise power spectral density (PSD) $\Phi_{pp}$ of these pressure fluctuations is obtained from the following equation:

\begin{equation}\label{-107}
    \Phi_{pp}(x,\omega) = \int_{-\infty}^{+\infty} R(x,\tau)e^{-i\omega t} \,d\tau
\end{equation}

Here, $R$ is the auto-correlation function defined as $R(x,\tau) = \int_o^\infty p^\prime(x,t)p^\prime(x,t+\tau) \,dt$, $\Phi_{pp}(x,\omega)$ is the PSD obtained from the fast Fourier transform of $R$, $\omega$ is the angular frequency and $i = \sqrt{-1}$. Several empirical models in the literature express the empirical correlation for $\Phi_{pp}$ in terms of the mean boundary layer parameters:

\begin{equation}\label{-108}
    \Phi_{pp}(\omega) = f(\omega,\delta,\delta^*,\theta,U_e,\rho,\nu,\tau_w,\partial_xp,c,\Pi)
\end{equation}

Where $\delta^* = \int_{0}^{\delta} (1-U(x,y)/U_e) \,dy$ is the displacement thickness, $\theta = \int_{0}^{\delta} U(x,y)/U_e(1-U(x,y)/U_e) \,dy$ represents the momentum thickness of the TBL, and $\tau_w$ is the wall shear stress at location $x$. $\rho$, $\nu$, and $c$ represent the density, dynamic viscosity, and speed of sound in the freestream respectively. A TBL typically consists of four distinct regimes in the wall-normal direction: viscous sublayer closer to the wall, intermediate buffer layer, logarithmic region influenced by large turbulent scales, and the wake region. The deviation of the velocity profile from the log law in the wake region is characterized by the wake strength parameter\cite{krug2017revisiting}, $\Pi$. The number of independent variables in Eq. \ref{-108} can be reduced using the Buckingham-Pi theorem. Following the work of Dominique \emph{et al.}\cite{Dominique2022}, Eq. ~\ref{-108} is rewritten in terms of the non-dimensionalize variables as:

\begin{equation}\label{-109}
    \widetilde{\Phi}_{pp} = f(\widetilde{\omega},\Delta,H,M,\Pi,C_f,R_T,\beta)
\end{equation}

Where,\\[-8pt]

Dimensionless PSD: $\widetilde{\Phi}_{pp} = (\Phi_{pp}U_e) / (\tau_w^2\delta) $ \\[-8pt]

Dimensionless angular frequency: $\widetilde{\omega} = (\omega\delta)/ U_e$ \\[-8pt]

Zagarola-Smits's parameter\cite{zagarola_smits_1998}: $\Delta = \delta /\delta^*$\\[-8pt]

Shape factor: $H = \delta^*/\theta$\\[-8pt]

Mach number: $M = U_e/c$\\[-8pt]

Wake strength parameter\cite{Coles1956}: $\Pi$\\[-8pt]

Friction coefficient: $C_f = \tau_w / (\rho U_e^2)$\\[-8pt]

Outer-to-inner-layer timescale ratio: $R_T = (\delta/U_e)/(\nu/U_\tau^2)$\\[-8pt]

Clauser parameter: $\beta = (\theta / (\rho U_\tau^2))\partial_xp$

\subsection{\label{sec1:subsec2}Goody's Model}

One of the earliest empirical models, the Chase-Howe model\cite{howe,Chase1980}, proposed a correlation of PSD ($\Phi_{pp}$) as a sole function of angular frequency $\omega$. It however fails to predict the high frequency $\widetilde{\omega}^{-5}$ drop observed in the wall pressure spectra of TBLs under ZPG. Goody\cite{Goody2004} proposed the following changes to the Chase-Howe model: (a) Since the largest coherent structures in TBL are of the order of boundary layer thickness, $\delta^*$ is replaced with $\delta$ based scaling (b) Ratio of outer-layer to inner-layer timescales, $R_T$ (a different form of Reynolds number) is incorporated into the formulation to control the extent of inertial subrange. Goody's model, given in Eq. ~\ref{-110}, accurately predicts ZPG turbulent boundary layers that are homogeneous in the spanwise direction. Hence it is considered as the baseline semi-empirical model although it does not account for pressure gradients.

\begin{equation}\label{-110}
    \frac{\Phi_{pp}U_e}{\tau_w^2\delta} = \frac{3(\omega\delta/U_e)^2}{[(\omega\delta/U_e)^{0.75}+0.5]^{3.7}+[1.1R_T^{-0.57}(\omega\delta/U_e)]^7}
\end{equation}

\subsection{\label{sec1:subsec3}Dominique's GEP Model}

Dominique \emph{et al.}\cite{Dominique2021} trained the gene expression programming algorithm on the datasets described in Section \ref{ch2:sec1} to come up with the following WPS model:

\begin{equation}\label{eq:6}
\widetilde{\Phi}_{pp}=\dfrac{\left(5.41+C_f\left(\beta+1\right)^{5.41}\right) \widetilde{\omega}}{\widetilde{\omega}^2+\widetilde{\omega}+(\beta+1) M+(\widetilde{\omega}+3.6) \dfrac{\widetilde{\omega}^{4.76}}{C_f R_T^{5.83}}}
\end{equation}

Although derived from scratch, Dominique's GEP model reasonably captures the frequency dependencies. Their model shows a $\widetilde\omega^1$ dependence at low frequencies, in contrast to the $\widetilde\omega^2$ dependence proposed by Kraichnan \cite{Kraichnan1956}. This is considered a strong trait for modeling realistic scenarios using GEP where the data does not always agree with the theoretical models. 

\section{\label{sec2}Methodology}
\subsection{\label{ch2:sec1}Data Collection}

The dataset used in the present work stems from the study by Dominique \emph{et al.}\cite{Dominique2022} (also referred to as the von Karman Institute (VKI) team henceforth). It consists of the variation of $\widetilde{\Phi}_{pp}$ with $\widetilde{\omega}$ at subsonic Mach numbers, in addition to the mean TBL characteristics (reported in Eq. ~\ref{-109}), collected from the following sources:

\begin{itemize}
    \item Experiments of Salze \emph{et al.}\cite{Salze2014} on a flat plate under ZPG, APG, and FPG.
    \item High-fidelity computations of Deuse and Sandberg\cite{Deuse2019}, Hao \emph{et al.}\cite{Wu2018} and, Christophe \emph{et al.}\cite{Christophe2009} on the configuration of the flow over a controlled diffusion (CD) airfoil, where the TBL spectra under APG and mild FPG are recorded at different stations.
\end{itemize}

\begin{figure}[!t]
\centering
\includegraphics[scale = 0.8]{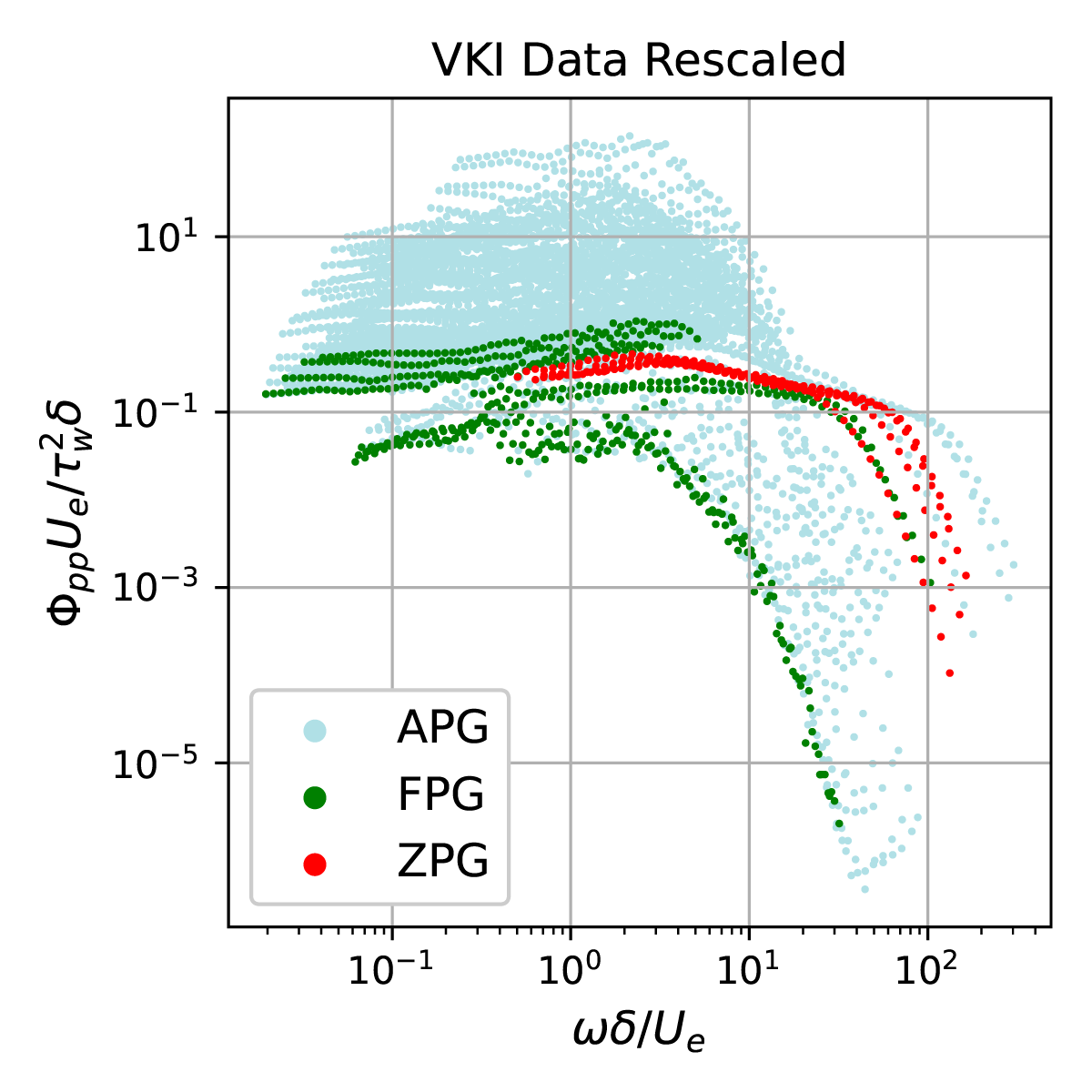}
\caption{\label{fig:vki_data_rescaled}Wall pressure spectra collected by the von Karman Institute(VKI) team rescaled to Goody's scales}
\centering
\end{figure}

\begin{figure}[!ht]
\centering
\includegraphics[scale = 0.8]{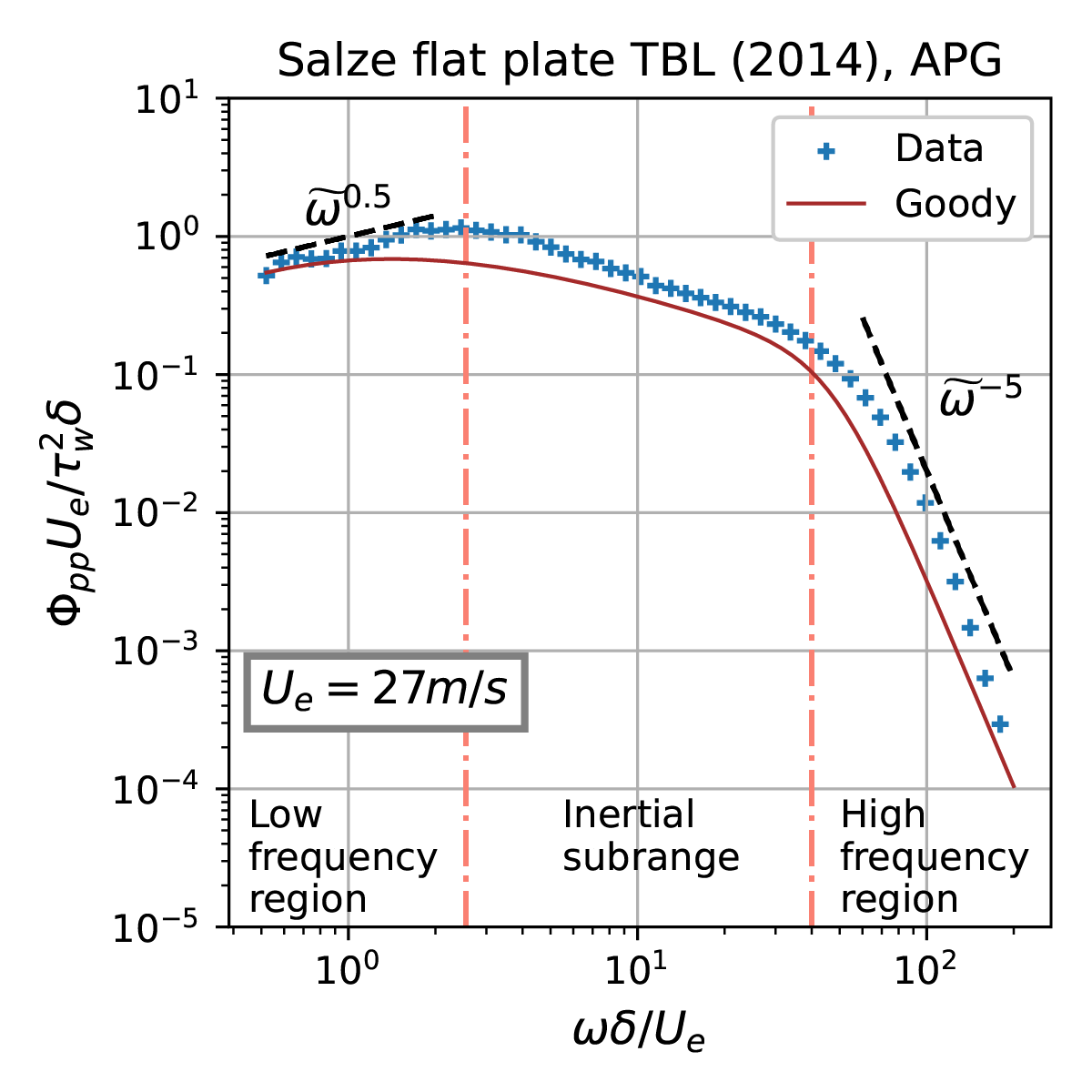}
\caption{\label{fig:wps_trends}Low and high-frequency trends in TBL wall pressure spectrum, black dashed lines as eye guides}
\centering
\end{figure}

Figure ~\ref{fig:vki_data_rescaled} plots all the 117 datasets, each of which has 500 logarithmically spaced points resampled from the experiments listed above. A machine learning algorithm is otherwise oblivious to the different experiments from which the dataset has been extracted. Rather, it solves the regression problem modeled using Eq. \ref{-109} treating the input as a set of independent data points with respective features and labels.

\begin{figure*}
\includegraphics[width = \linewidth]{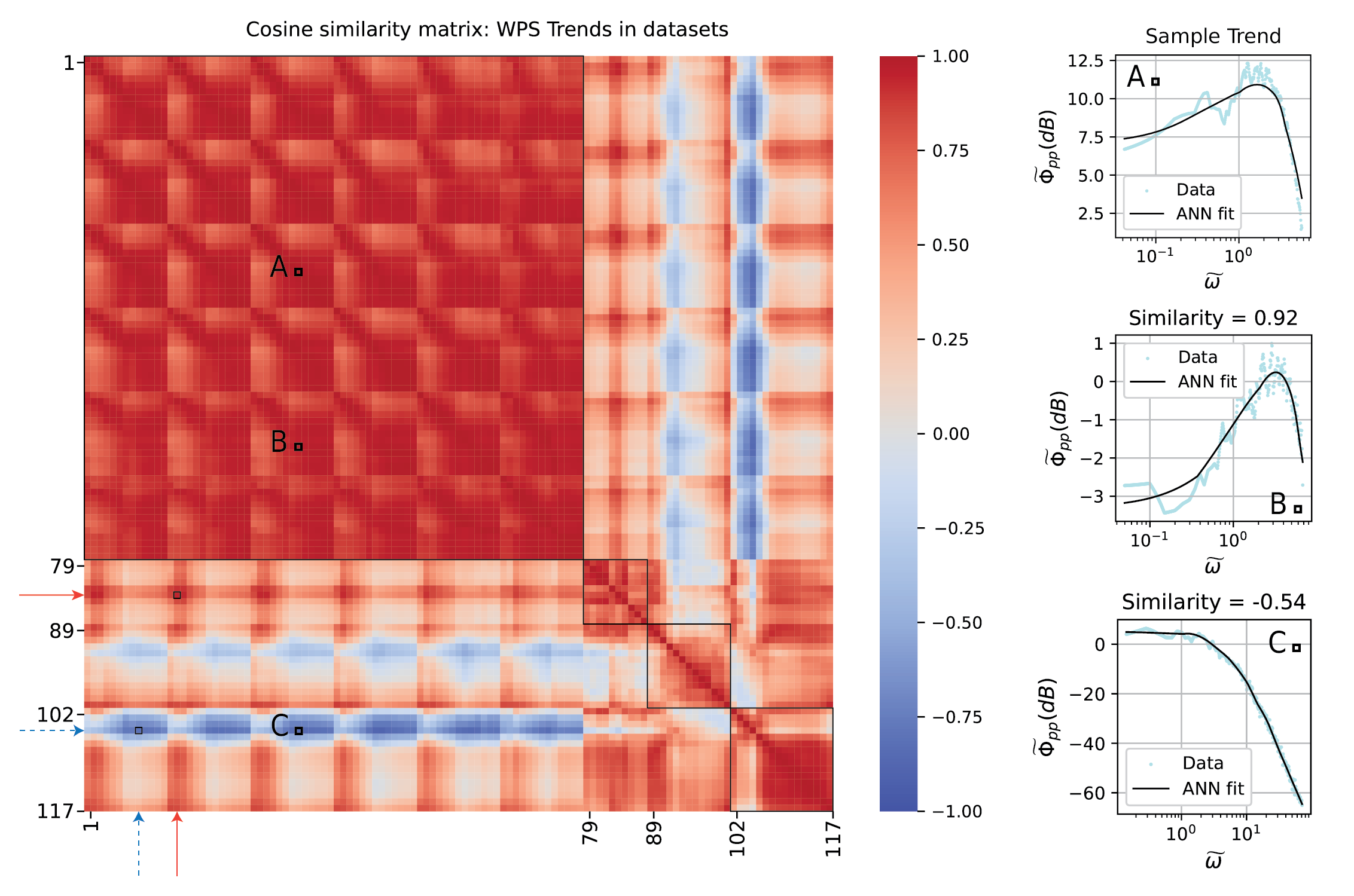}
\caption{\label{fig:cosine_similarity}Cosine similarity matrix of WPS trends observed in collected datasets.\\
Christophe:1-78, Salze:79-88, Deuse: 89-101: Hao: 102-117}
\centering
\end{figure*}

\subsection{\label{ch2:sec2}Data analysis}

A closer look at Fig.~\ref{fig:vki_data_rescaled} shows the dominance of the APG dataset over the FPG and ZPG. It is essential to design ML algorithms to handle this skewness, else it treats the underrepresented cases as outliers and the model predictions suffer. A common practice to balance out skewed databases is to identify similarity clusters and assign the respective weights. One of the widely used methods involves clustering based on the Euclidean distance, $d = \sqrt{\sum_{i=1}^n (P_i - Q_i)^2}$ between two vectors (say $\overline{P}$ and $\overline{Q}$) which represent the independent flow parameters listed in Eq. \ref{-108}.

In the current study, we demonstrate the similarity by comparing the trends of $\widetilde{\Phi}_{pp}$ vs $\widetilde{\omega}$ across datasets. A typical WPS, shown in Fig. \ref{fig:wps_trends}, comprises three distinct regimes: a low-frequency region where $\widetilde{\Phi}_{pp}$ rises monotonically with $\widetilde{\omega}$ to reach its peak value, an inertial subrange at mid-frequencies (typically noticeable at high Reynolds numbers) and a $\widetilde{\omega}^{-5}$ roll-off at higher frequencies due to dissipation. However, not all the datasets in Fig.~\ref{fig:vki_data_rescaled} show this exact trend. The range of frequencies encompassing each of the regimes also differs across datasets. Hence, the following slope vector $\overline{m}$ is extracted for each dataset of length $n$:

\begin{equation}\label{-112}
    m_i = \frac{\widetilde{\Phi}_{pp,i+1}(dB) - \widetilde{\Phi}_{pp,i}(dB)}{\widetilde{\omega}_{i+1}-\widetilde{\omega}_{i}}, i = [1,n-1]
\end{equation}

It should be noted that the raw WPS dataset is noisy and hence should be smoothed out to extract meaningful trends. The present work used the predictions of an ANN model for the same. Since all datasets have an equal number of points, $\overline{m}$ also captures the extent of respective regimes. We can now compute the cosine similarity, which results in a value between [-1,1], to quantify the closeness between two slope vectors as follows:

\begin{equation}\label{-113}
    CS = cos(\overline{m}_i,\overline{m}_j) = \frac{\overline{m}_i\cdot\overline{m}_j}{\lvert\lvert\overline{m}_i\rvert\rvert\times \lvert\lvert\overline{m}_j\rvert\rvert}, (i,j) = [1,117]
\end{equation}

In a 3D space, $CS = 1$ represents similar vectors, $CS = 0$ indicates orthogonal vectors and $CS = -1$ represents vectors pointing in opposite directions. In a multidimensional space such as ours, cosine similarity is merely representative of how similar trends are to each other; from identical to very dissimilar respectively. Figure ~\ref{fig:cosine_similarity} plots the heatmap of the resulting cosine similarity matrix. $CS \approx 1$ within the boxes aligned along the diagonal indicate that the trends within a given experiment are similar. Bands of $CS \approx 1$ (highlighted using solid arrows) represent similar WPS trends observed across the experiments with different flow conditions. In contrast, bands of $CS \approx -1$ (highlighted using dashed arrows) represent dissimilar trends. For example, the inset plots in Fig. ~\ref{fig:cosine_similarity} illustrate the similarity of WPS extracted at points A, B, and C. It is apparent that the WPS at B and A are similar with a $CS \approx 0.92$ while the WPS at C is dissimilar to that of A with a $CS \approx -0.54$. This similarity matrix helps in identifying clusters that can be used to assign weights to the datasets and potentially improve model predictions. This aspect is further discussed in Section ~\ref{ch3:sec1}.


\subsection{\label{ch2:sec3}Artificial neural networks}

\begin{figure}
\centering
\includegraphics[width =\linewidth]{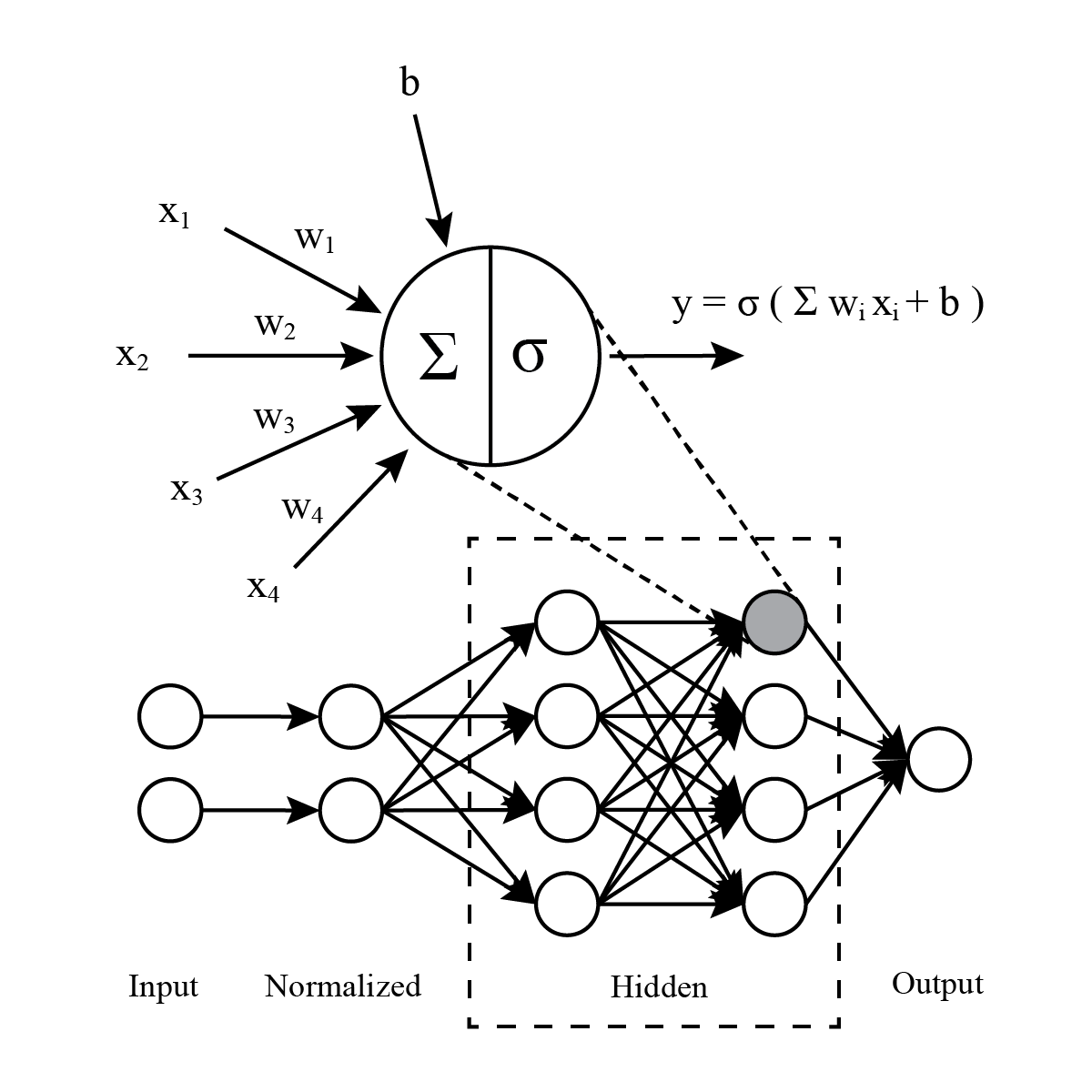}
\caption{\label{fig:neural_network}Example of an Artificial neural network build}
\end{figure}

Artificial neural network (ANNs) is a powerful function approximator that is commonly used to fit non-linear data using supervised learning. A brief description of ANN is given here and readers can refer to the works of Goodfellow \emph{et al.}\cite{MIT} for further details. Figure ~\ref{fig:neural_network} illustrates a typical ANN architecture comprising an input layer (which can be normalized for improved scaling of data), several hidden layers (to accurately capture the complexity of the non-linear function that fits the data), and an output layer. The hidden layers consist of neurons. Each neuron generates a weighted sum $\Sigma$ from its inputs ($\overline{w} \cdot \overline{x}$) with respective biases $b$. A predefined activation function ($\sigma$, usually a non-linear function) operates on this weighted sum $\Sigma$ to produce an output $y$. The bias is a constant value that is added to the weighted sum of the inputs to shift the activation function. It is mathematically expressed as $y = \sigma\Sigma = \sigma(\overline{w} \cdot \overline{x} + b)$. The error between the expected output and the prediction is minimized by computing the gradient of the objective function with respect to the weights and biases of each neuron through backpropagation\cite{rumelhart1986learning}. This helps in adjusting the weights and biases of the individual neurons, resulting in a potentially improved model.

The current study considers the data used by the VKI team and hence the optimal hyperparameters from their study have been used to build the ANN model. The neural network has three hidden layers of ten neurons each in a fully-connected feed-forward fashion. These hidden layers are preceded by a normalization layer. In this layer, the input vector $\mathit{\overline{x}} = [\widetilde{\omega},\Delta,\mathit{H},\mathit{Ma},\Pi,\mathit{C_f},\mathit{R_T},\beta]$ from Eq.~\ref{-109} is normalized using the mean and standard deviation of each feature to accelerate the learning process\cite{Ba2016}. The output layer of the model is made up of a single neuron that predicts the value, $\mathit{y} = 10log_{10}(\Phi_{pp}U_e/\tau_w^2\delta)$. The training is carried out using Nadam optimizer\cite{Dozat2016} and Selu activation function\cite{Klambauer2017} at a learning rate of 0.0001 with a batch size of 32 randomly chosen samples. The dataset is randomly split into training data and validation data with an 80:20 split respectively. Logarithmic mean squared error ($lMSE$), defined in Eq. \ref{eq:-1}, is used as the objective function. Training of the ANN is driven by the training loss while the early stopping is identified by monitoring the validation loss.

\begin{equation}\label{eq:-1}
\mathit{lMSE} = \frac{1}{\mathit{N}}\sum\limits_{i=1}^N\mathit{W_i}(  10log_{10}(y_i^{expected}) - 10log_{10}(y_i^{predicted}))^2
\end{equation}

\subsection{\label{ch2:sec4}Gene expression programming (GEP)}

\begin{figure*}
\centering
\includegraphics[width = 0.98\linewidth]{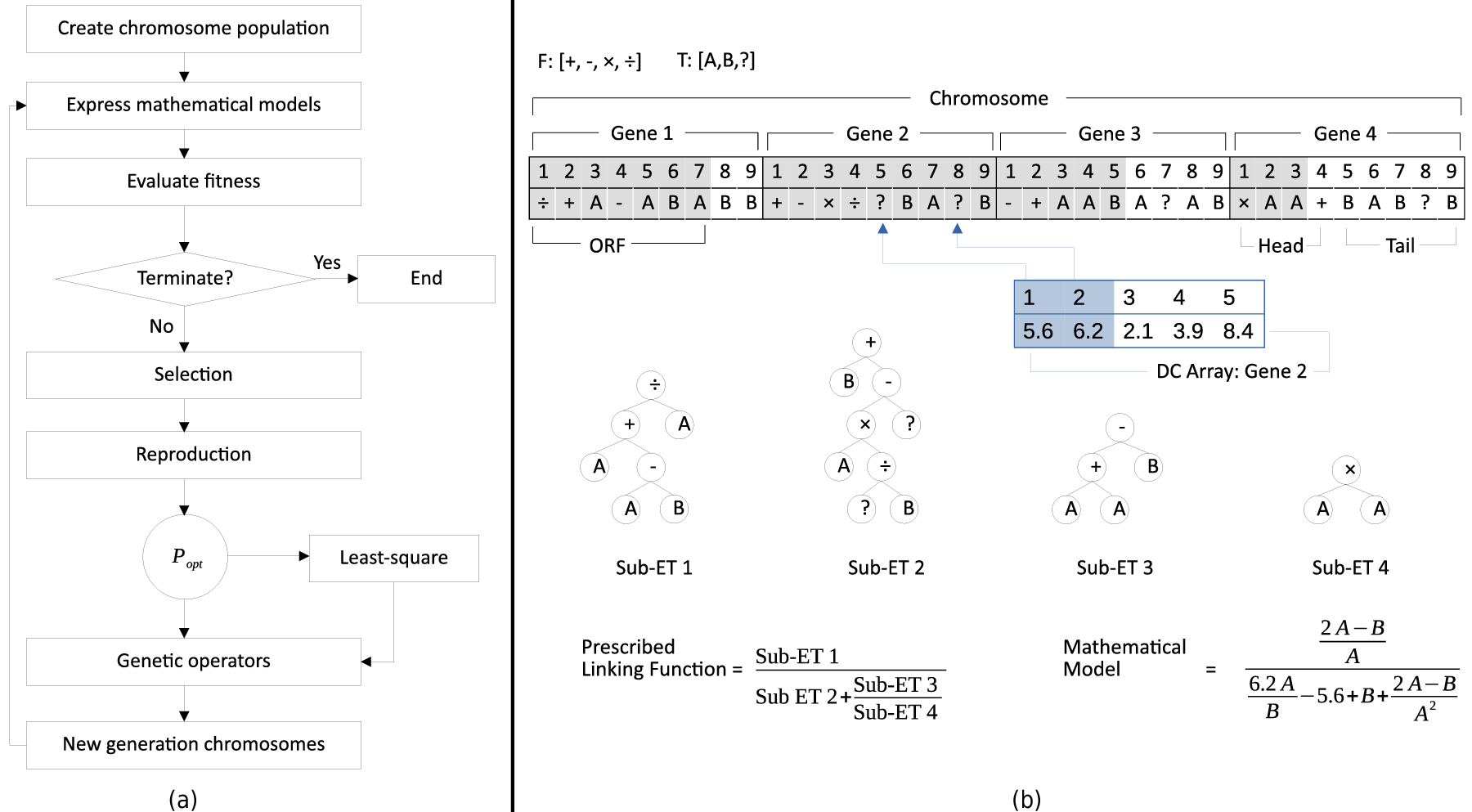}
\caption{\label{fig:gene}(a) GEP algorithm flowchart with least-square based local optimization. (b) Entities involved in GEP}
\centering
\end{figure*}

\begin{table*}
\begin{ruledtabular}
\centering
\caption{\label{tab:table1}GEP Terminology}
\renewcommand{\arraystretch}{1.2}
\begin{tabular}{p{0.25\linewidth}|p{0.7\linewidth}}
\textbf{Term}                           & \textbf{Description}                                                                                                                                                          \\ \hline
Functional set (F)             & Set of operators that a GEP inferred mathematical expression can be built out of.                                                                                    \\ \hline
Terminal set (T)               & Set of variables that a GEP inferred mathematical expression can be built out of.                                                                                    \\ \hline
Gene                           & Fixed length string that is a combination of members from the functional and the terminal set.                                                                       \\ \hline
Head                           & Part of the gene that is a combination of entities from the functional set and the terminal set.                                                                     \\ \hline
Tail                           & Part of the gene made exclusively out of the terminal set to ensure that enough terminals are available to result in a valid mathematical expression.                \\ \hline
Open reading frame (ORF)       & Part of the gene that can be read into an expression tree.                                                                                                           \\\hline
Expression tree (ET)           & Plane tree structure with each node as a member of the functional set and each leaf as a member of the terminal set that can be read into a mathematical expression. \\\hline
Chromosome/Individual          & Fixed length string that consists multiple genes.                                                                                                                    \\\hline
Sub Expression Tree (Sub - ET) & Expression tree that is derived from a gene present in a chromosome.                                                                                                 \\\hline
Prescribed Linking function    & Function prescribed at the start of GEP training that combines Sub - ETs to produce the final mathematical expression from a chromosome.                             \\\hline
Fitness                        & Performance of GEP individuals quantified through objective functions.                                                                                               \\\hline
Selection                      & Process of selecting parents for the next generation from a pool of individuals while discarding the rest.                                                           \\\hline
Reproduction                   & Process of filling up the population back to its original volume by deriving daughters from parent individuals.                                                      \\\hline
Genetic operators              & Operators that genetically modify chromosomes.                                                                                                                       \\\hline
Evolution                      & Process of performing selection and reproduction on a population while retaining the desired genetic characteristics identified through fitness.                     \\\hline
Generation                     & Pool of individuals that undergoes fitness evaluation at any point of time through GEP evolution.                                                                   
\end{tabular}
\end{ruledtabular}
\end{table*}

Like a genetic algorithm GEP discovers mathematical expressions to best represent a dataset. We present a brief summary of the method here and the readers can refer to the article by Candida F.\cite{gep} for a detailed discussion on GEP. Figure \ref{fig:gene} and Table \ref{tab:table1} illustrate the mainstream GEP algorithm, the entities, and the operations involved. The algorithm starts with constructing several fixed-length strings called genes using combinations of the functions and terminals. Part of such a gene (ORF) is interpreted as an expression tree (ET) which is a graphical representation of a mathematical expression. Chromosomes (strings with multiple genes) with linking functions are used to improve upon the complexity captured by the mathematical expression derived from a single gene. The fitness of these chromosomes is then evaluated based on their ability to match the predictions of a given dataset. Chromosomes with desired fitness are retained, reproduced, and modified through genetic operators to result in different expressions altogether. Well-performing combinations are thus retained and mutated over multiple generations until the one with desired fitness is discovered.

\subsubsection*{\label{ch2:sec4:subsec1:subsubsection:8}Numerical constants and the power terminal}

In mathematical expressions, there are often numerical constants associated with the variables which are considered as terminals. GEP computes these constants using methods like Random Numerical Constants (RNC) where an extra terminal, say `?', is introduced in the terminal set. Each gene is additionally assigned an array of numerical constants called a Dc array. These numerical constants sequentially replace the `?' when encountered in an ORF. The Dc array can be evolved in a similar manner as the gene sequence.

Dominique \emph{et al.}\cite{Dominique2021} implemented an additional `pow' terminal to raise a variable by a constant power. It works in a similar fashion as the RNC terminal and selects a power constant from the Dc array whenever this terminal is encountered in ORF. Unlike traditional GEP, this allows for introducing variables raised with fractional powers. This terminal also reduces the length of a gene significantly while achieving the same degree of complexity in the expression. Randomly selecting numerical powers, however, results in high divergence in population. This is controlled by introducing an additional local optimizer as shown in Figure \ref{fig:gene}(a). The optimizer, although computationally expensive, searches for the best exponent of a variable within a predefined range using the least squares optimization technique. The algorithm is thus modified with two new user-defined hyperparameters: (a) $P_{opt}$, which is a probability of selecting an individual for the local optimizer to go through the least squares optimization and (b) the optimization period, which determines the number of generations after with this optimization is allowed.

A modified version of the Python library, geppy, with the power terminal capability by the VKI team, has been used in the current work. The results discussed in Section \ref{sec3} use this version of the geppy package. Additional modifications have been incorporated into this version to accelerate the convergence of GEP through guided search as will be discussed in Section \ref{ch4}.

\section{\label{sec3}Results}
\subsection{\label{ch2:sec4:subsec2}Hyperparameter study: GEP}

The GEP algorithm by its very nature is unsupervised and hence extremely difficult to converge. The extent of selection and mutation, the two key processes that outline the path taken by the algorithm to converge, is an optimization problem in itself. Head length is another parameter that determines the complexity of solutions. A highly complex solution may result in a lower error but at the same time suffer from overfitting the data; often indicated by poor predictions on the unseen data. It hence becomes necessary to optimize the hyperparameters that lead to a solution of the right complexity and the desired fit within a reasonable time. For simplicity, a dataset derived from Goody's model (Eq. ~\ref{-110}) is considered. This allows us to verify if the GEP is able to find a known original expression from which the input dataset is derived.



\begin{figure*}
\centering
\includegraphics[width = \linewidth]
{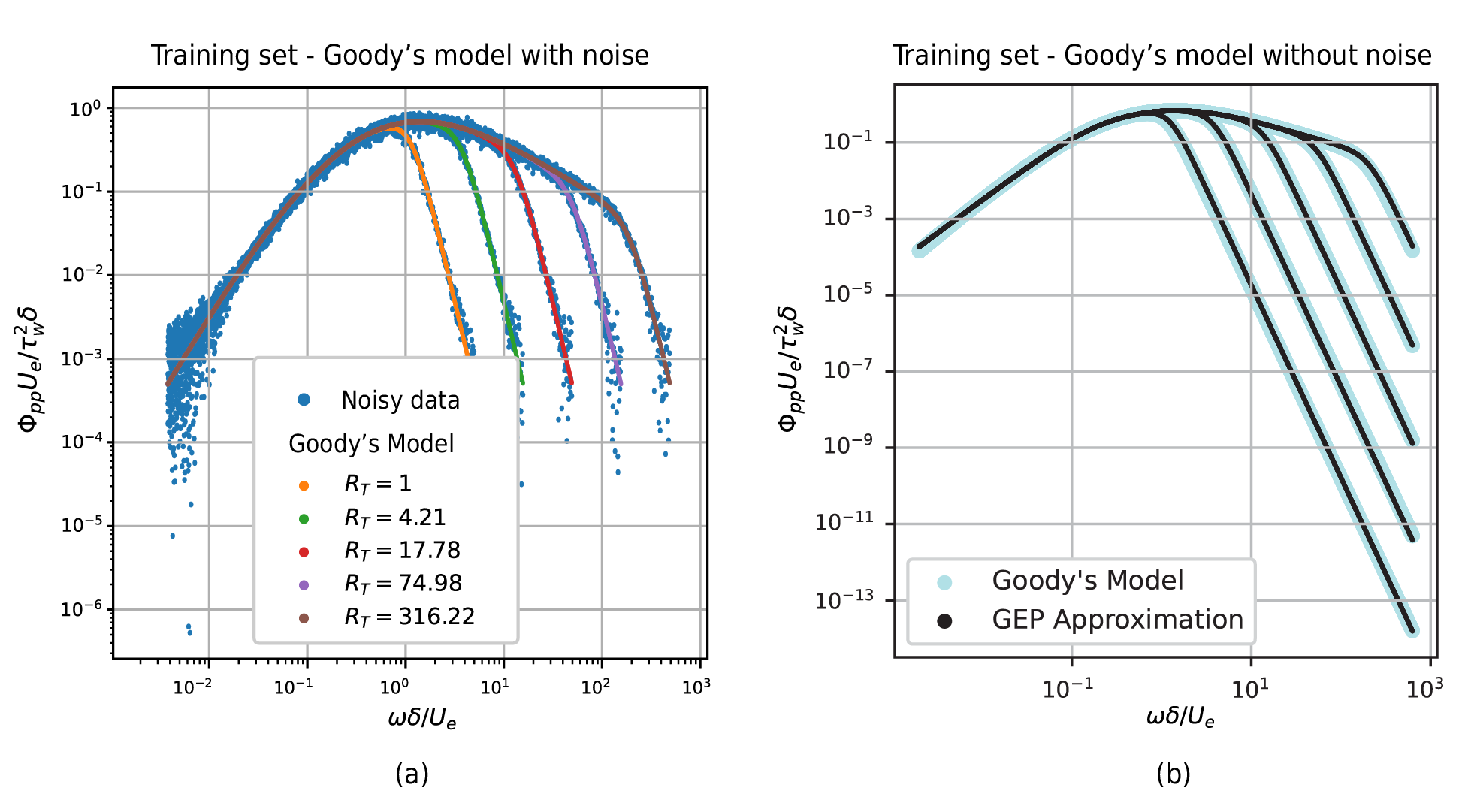}
\caption{\label{fig:training_GEP} Training dataset for hyperparameter study of GEP (a) With added noise (b) Without any noise along with a typical GEP solution observed at h = 4, 10001 generations and, $P_{opt} = 0.1$}
\centering
\end{figure*}

As discussed in Section \ref{sec1:subsec2}, Goody’s model is dependent on two variables, $\widetilde{\omega}$ and $R_T$. It has a $R_T^{4}$ dependence, an $\widetilde{\omega}^{2}$ dependence at low frequencies and, an $\widetilde{\omega}^{-5}$ dependence at high frequencies \cite{Goody2004}. These features were specifically looked at in the GEP solutions to quantify their closeness to Goody’s model. 
The training dataset constituted of 5 points of $R_T$ logarithmically spaced between [1, $10^{2.5}]$ and 2000 points of $f=\omega/2\pi$ logarithmically spaced between [$10^{-3.5}$, $10^{2}$] for each $R_T$ considered. The resulting 10000 points of $\widetilde{\Phi}_{pp}$ thus obtained from these values were further introduced with Gaussian noise to represent the typical static and stochastic uncertainties that are found in the experimental data. A lower threshold of 0.0005 on $\widetilde{\Phi}$ is set to avoid additional noise that would otherwise make GEP impossible to converge. The training dataset thus obtained is shown in Fig.~\ref{fig:training_GEP} (a) and Fig.~\ref{fig:training_GEP} (b) shows the corresponding dataset without noise.

\begin{figure*}
\centering
\includegraphics[width=0.8\linewidth]{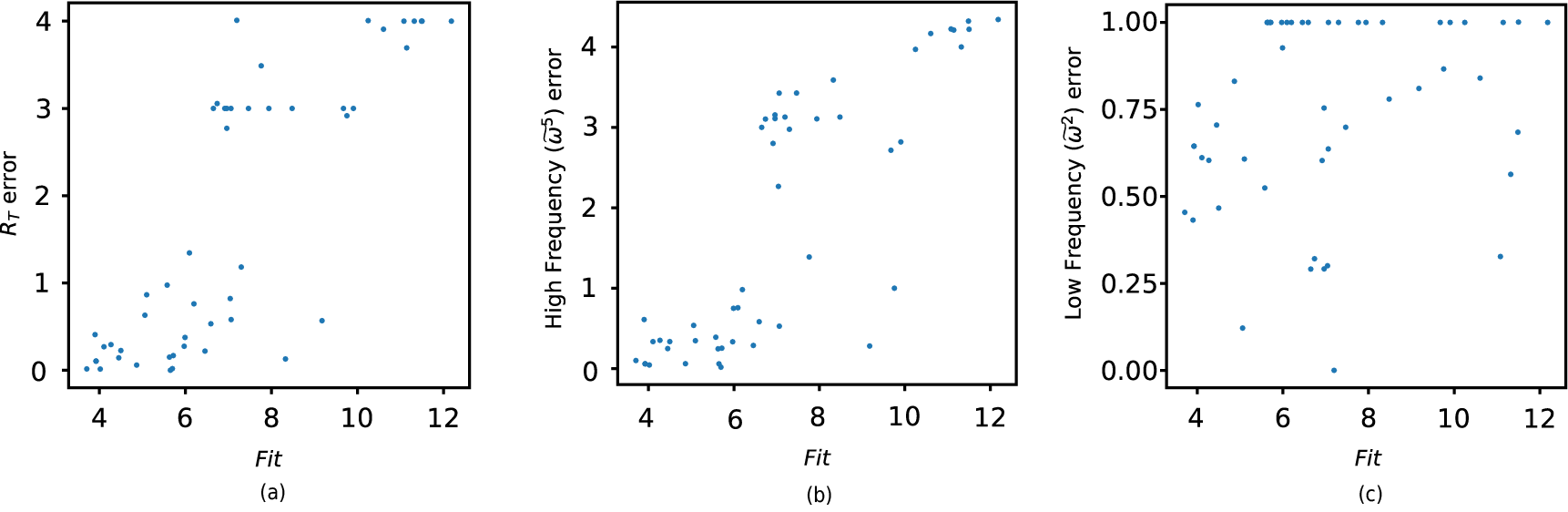}
\caption{\label{fig:exponent_error}Exponent error observed in GEP solution while predicting the Goody’s model}
\centering
\end{figure*}

The objective function to be minimized comprises a weighted contribution of $lMSE$ and $MSE$, which is given as:

\begin{equation} \label{eq:2}
    \mathit{Fit} = 100\times\sqrt{\alpha\frac{\mathit{lMSE}}{A_{log}^2} + (1-\alpha)\frac{\mathit{MSE}}{A_{lin}^2}} 
\end{equation}

where,

\begin{equation}\label{eq:3}
    \mathit{MSE} = \frac{1}{N}\sum_{i=1}^NW_i(y_i^{expected}-y_i^{predicted})^2
\end{equation}

Here, $A_{log}$ and $A_{lin}$ represent the maximum amplitude of training points in the respective scales. $\alpha$ is a weight assigned to the objectives, which is set to 0.5 thus giving equal weight to both contributions. The idea to have such a multi-objective function is derived from the works of Dominique \emph{et al}. \cite{Dominique2021}. They postulated that while the $MSE$ penalizes the error in the amplitude of the spectrum, the $lMSE$ captures error in the trends at low and high frequencies effectively.

\begin{figure*}[]
\centering
\includegraphics[width = \linewidth]{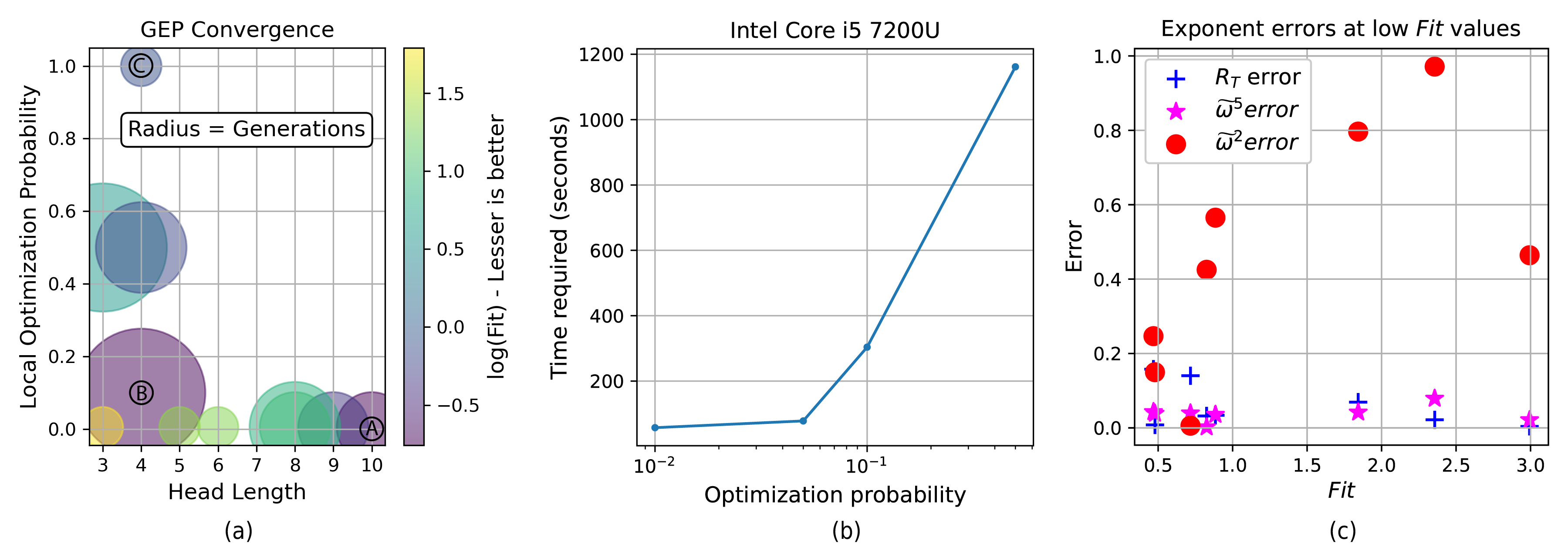}
\caption{\label{fig:scatter_fit}(a) $\mathit{Fit}$ (log($\mathit{Fit}$) as colorbar) as a function of head length (x-axis), local optimization probability (y-axis), and number of generations (area of circle). (b) Effect of optimization probability on the computational time required to run the algorithm through 200 generations for h = 4 (c) Predictable drop observed in low-frequency error at low $\mathit{Fit}$ values}
\centering
\end{figure*}

A set of experiments were conducted for 100 individuals of different head lengths [3,4,5,6], over 1001 generations, and with optimization probability 0.005 (every generation is being optimized, but very few individuals are optimized in each generation). Figure ~\ref{fig:exponent_error} plots the errors in exponent predictions (which is defined as $\lvert \mathit{Exponent_{expected}}-\mathit{Exponent_{observed}} \rvert$) with $\mathit{Fit}$. It is evident that the $\mathit{R_T}$ and high-frequency errors tend to drop with a decrease in $\mathit{Fit}$, implying a closer agreement with Goody’s model. In almost all the cases, $\mathit{R_T}$ error has dropped below 0.5 for $\mathit{Fit} < 5$  and below 1.0 for $\mathit{Fit} < 6$. Similar behavior is observed for high-frequency error where the error drops below 1 for $\mathit{Fit} < 6$. However, it is evident that the low frequency $\widetilde{\omega}^2$ errors are quite unpredictable, although the errors are $\leq 1.0$ for all the cases. This unpredictable behavior is attributed to the high noise at low frequencies, a feature of input data that is observable in Fig.~\ref{fig:training_GEP} (a). As pointed out by the Dominique $et$ $al$ \cite{Dominique2021}, $\mathit{Fit}$ by definition is a cumulative error criterion. Hence, a lower value of $Fit$ for a solution does not guarantee the local closeness to the parent function (Goody’s model) in small individual subdomains.

The value of $Fit$ with the noisy data reported in Fig. ~\ref{fig:exponent_error} is $\approx \geq 4$. Subsequent studies were hence carried out with clean data (see Fig.~\ref{fig:training_GEP} (b)) to determine if the GEP predictions improved. Figure ~\ref{fig:scatter_fit} (a), shows a scatter plot with varying radius and color to demonstrate the effect of hyperparameter setup (head length, local optimization probability, generations) on the model performance ($log(Fit)$). Here, the color and radius of the circle represent the $log(Fit)$ and the number of generations evolved in each trial respectively. Following are the key inferences that can be drawn from the figure:

\begin{itemize}
    \item As expected, improved values of $Fit$ (below 1.0) can be achieved with clean data when compared to noisy data. 
    \item $\mathit{Fit}$ drops with increasing head length, although with larger head lengths there is a risk of making the model too complex. Note that the $Fit$ values were still higher with increasing head lengths (up to 10) with 1001 generations (not shown in the plot). However, with further optimization, $\mathit{Fit}$ is observed to drop below 1 for h = 10 and 3001 generations (refer to marker A in Figure ~\ref{fig:scatter_fit} (a)). This is because a higher number of generations allow for more mutations, thereby improving the $Fit$. Decreasing the head length also demands a larger number of generations which translates to more time spent in model evolution. Local optimization can circumvent this issue by minimizing the divergence within the population generated by the added power feature using the local least squares optimizer discussed in Section \ref{ch2:sec4:subsec1:subsubsection:8}.
    \item Increasing the local optimization probability improves the $\mathit{Fit}$ even at lower head lengths with less number of generations. Nevertheless, this considerably increases the time per generation as illustrated in Fig.~\ref{fig:scatter_fit} (b). Figure ~\ref{fig:scatter_fit} (a) shows that increasing the optimization probability from 0.005 to 0.1 drops $\mathit{Fit}$ below 1 for a head length of 4 trained over 10001 generations (refer to marker B in Figure ~\ref{fig:scatter_fit} (a)). The fact that a solution at the complexity of h = 4 produced comparable predictions to the ones observed for h = 10 shows the potential of the local optimizer. Further increasing the optimization probability to 1, dropped the required number of generations to 1001 resulting in a $\mathit{Fit}$ below 1, (refer to marker C in Figure ~\ref{fig:scatter_fit} (a)).
    \item Higher number of generations results in a lower $Fit$ even at lower head lengths and optimization probabilities (refer to marker B in Figure ~\ref{fig:scatter_fit} (a)). 
\end{itemize}


All these insights lead to the conclusion that a user should prioritize mutations by evolving GEP over a greater number of generations. The subsequent diverging behavior due to mutations, if any, can be controlled with increased use of the local optimization loop. If both these solutions are ineffective, the user can consider increasing the head length as the GEP expressions with a smaller head length may not be complex enough to capture the features of the data. Figure ~\ref{fig:scatter_fit} (c), shows the encouraging trend that the proposed setup that reduces the $Fit$ error also reduces the low frequency $\widetilde{\omega}^2$ error which was found to be unpredictable earlier. Figure ~\ref{fig:training_GEP} (b) illustrates the typical curve fit obtained with GEP (with $h=4$, $P_{opt}=0.05$ and 10000 generations) where the $\mathit{Fit}$ drops below 1.

Due to its unsupervised evolutionary nature, GEP can produce ill-performing solutions and there is always a possibility of the algorithm not converging to the desired accuracy. The aforementioned hyperparameter tuning strategies aim to minimize such issues with convergence. As will be demonstrated in Section \ref{ch4}, unlike ANN, data of reasonable quality (with minimal noise and outliers) should also be provided to ensure convergence of GEP.

\subsection{\label{ch3:sec1}Weights and Objective functions}

It is established in Section \ref{sec2} that the database used in this study derives from different experiments with their own set of flow conditions. Table \ref{tab:table4} consolidates different experiments and the corresponding number of datasets, $N$. It is worth noting that the datasets collected within the same experiment can be mutually similar and differ greatly from those of other experiments which have different flow conditions. If every dataset is given equal weight/importance in the overall objective function, the model will be biased towards the over-represented data (for example, the data from Christophe). The rest of the minority datasets have little effect on the direction of the solution and suffer from poor model predictions even if the datasets are just as important\cite{Cui2019}. One way to mitigate this problem is through weighting datasets to balance the objective function. The selection of weights is a user decision and an optimization problem in itself which is beyond the scope of this work. To illustrate the effect of a weighted objective function on the ANN and GEP predictions, a rather simple weighting strategy is considered where the weight is estimated as the ratio of the number of datasets in the least represented experiments (the data from Salze) to the experiment considered i.e. $W = N_{Salze}/N_{Experiment}$. The weights assigned this way are listed in Table ~\ref{tab:table4}. 

\begin{table}[!b]
\centering
\caption{\label{tab:table4}Weights assigned to different experiments}
\begin{ruledtabular}
\begin{tabular}{lll}
Experiment & {N(Datasets)} & Weight \\ \hline
Salze      & 10          & 1.00   \\ \hline
Deuse      & 13          & 0.77   \\ \hline
Hao        & 16         & 0.62   \\ \hline
Christophe & 78        & 0.13   \\ \hline
Total      & 117               
\end{tabular}
\end{ruledtabular}
\end{table}

\begin{figure}[!t]
\centering
\includegraphics[scale = 0.65]{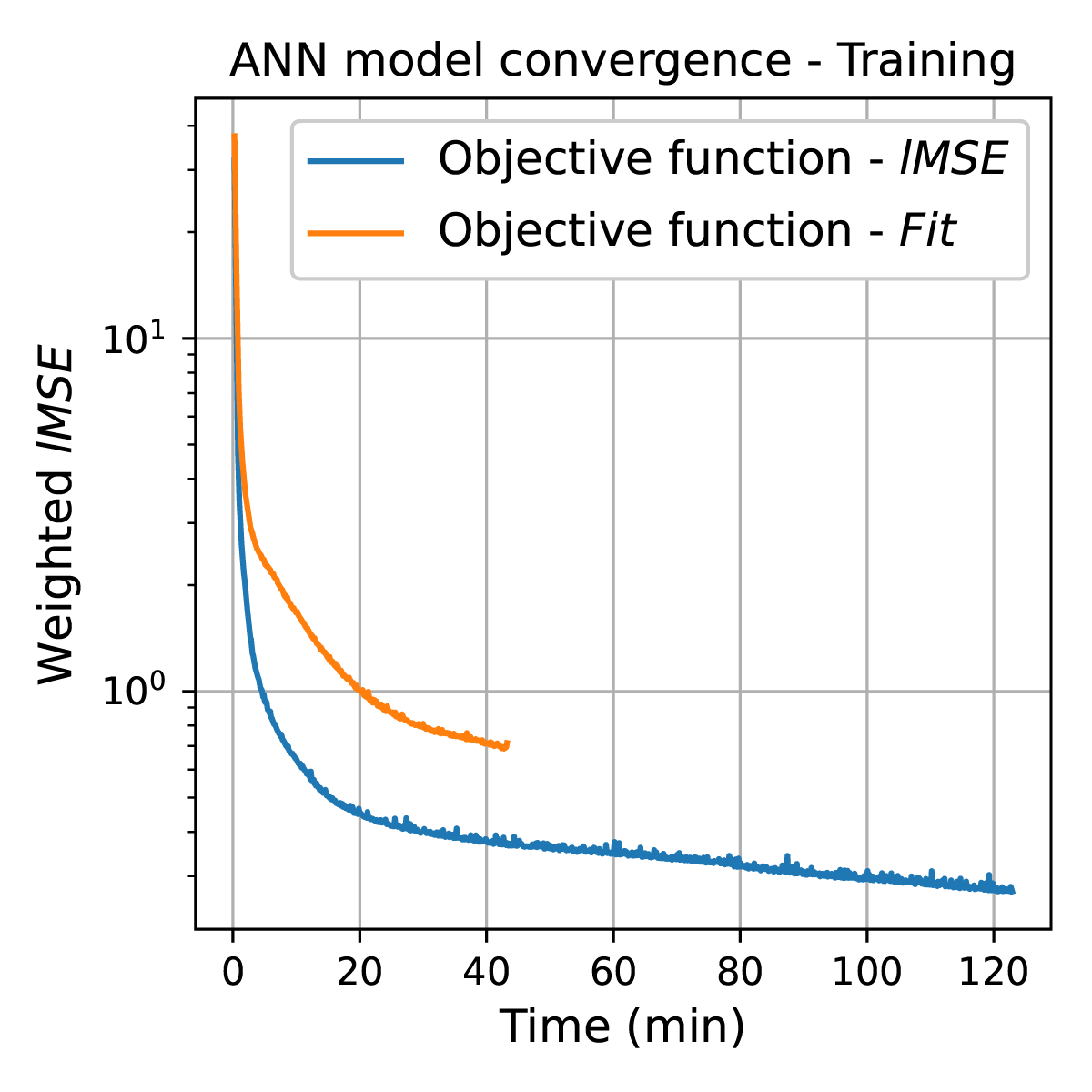}
\caption{\label{fig:ann_model_convergence_training}Example of the evolution of model performance in ANN training with (a) weighted $lMSE$ and (b) weighted $Fit$ as objective functions respectively.}
\centering
\end{figure}

\begin{table*}
\centering
\caption{\label{tab:table51} Comparison of $lMSE$s in predictions across experiments}
\begin{ruledtabular}
\begin{tabular}{lll|lll|lll}
\multirow{2}{*}{Experiment} & \multirow{2}{*}{\begin{tabular}[c]{@{}l@{}}N\\ (Datasets)\end{tabular}} & \multirow{2}{*}{Weight} & \multicolumn{3}{l|}{ANN ($lMSE$ error)}                                                                                                                                                                                    & \multicolumn{3}{l}{GEP ($lMSE$ error)}                                                                                                                                                                                     \\ \cline{4-9} 
                            &                                                                         &                         & \begin{tabular}[c]{@{}l@{}}$lMSE$\\ (weighted)\\ training\end{tabular} & \begin{tabular}[c]{@{}l@{}}$lMSE$\\ (unweighted)\\ training\end{tabular} & \begin{tabular}[c]{@{}l@{}}\%\\ change\\ w.r.t\\ weighted\end{tabular} & \begin{tabular}[c]{@{}l@{}}$lMSE$\\ (weighted)\\ training\end{tabular} & \begin{tabular}[c]{@{}l@{}}$lMSE$\\ (unweighted)\\ training\end{tabular} & \begin{tabular}[c]{@{}l@{}}\%\\ change\\ w.r.t\\ weighted\end{tabular} \\ \hline
Salze                       & 10                                                                      & 1.00                     & 0.24                                                                   & 0.33                                                                     & +37.50                                                                 & 6.98                                                                   & 9.43                                                                     & +35.10                                                                 \\ \hline
Christophe                  & 78                                                                      & 0.13                    & 0.39                                                                   & 0.23                                                                     & -41.03                                                                 & 7.41                                                                   & 3.81                                                                     & -48.58                                                                
\end{tabular}
\end{ruledtabular}
\end{table*}

Both GEP and ANN were trained on the same input data with weighted and unweighted objective functions. Technically training can be done indefinitely and the loss calculated by the objective function may continue to drop till it reaches a minimum. However, after a steep drop of the $lMSE$, Fig.~\ref{fig:ann_model_convergence_training} shows a marginal improvement in an ANN model performance with a further increase in epochs. For practical purposes, an early stopping criterion is therefore used to halt the training process. This can be done in a number of ways, say by setting an upper bound on epochs or by setting a lower bound on the loss calculated by the objective function. Here for this test, $20\%$ of the randomly sampled training data is set aside as validation data. The calculated loss against validation data was then monitored and training was stopped if there is no improvement for at least 20 epochs, followed by restoring ANN weights (not to be confused with database weights) to the ones observed for the epoch with the least validation loss. Figure ~\ref{fig:ann_model_convergence_training} also shows that the rate of accuracy gain (indicated by the rate of drop of error reported as weighted $lMSE$) is higher for the ANN models trained with $lMSE$ as the objective function than the ones trained with $Fit$.

\begin{figure*}
\centering
\includegraphics[scale = 1]{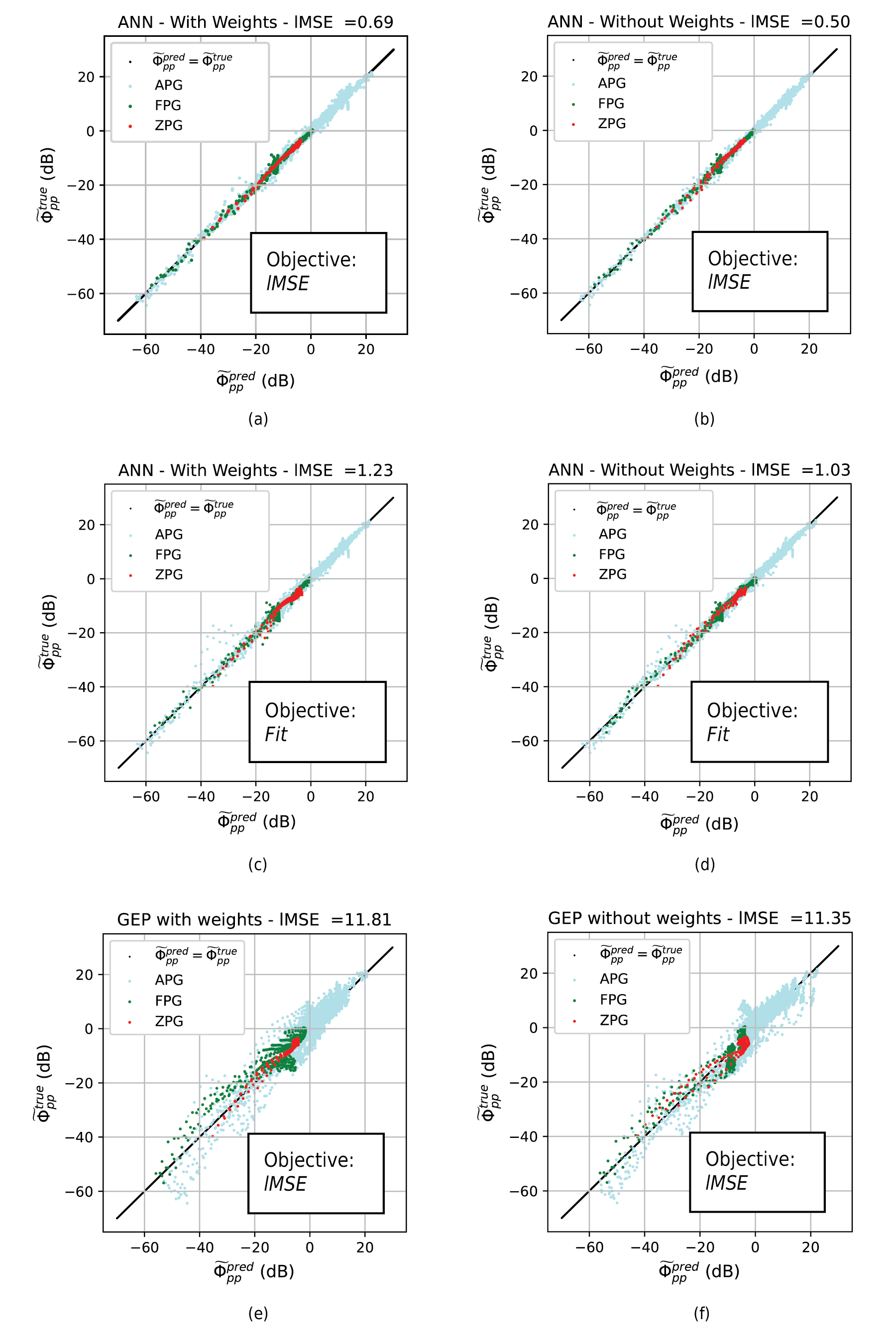}
\caption{\label{fig:ann_with_without_weights_lmse_fit_pred_low}Predictions of ANN model trained with weighted (a) and unweighted (b) $lMSE$ as the objective function; weighted (c) and unweighted (d) $Fit$ as the objective function. Predictions of GEP models, Eq.~\ref{eq:4} and Eq.~\ref{eq:5}, trained with weighted, (e), and unweighted $lMSE$, (f), respectively as the objective functions. (Every tenth point is reported for plotting purposes)}
\centering
\end{figure*}

\begin{figure*}
\centering
\includegraphics[scale = 1]{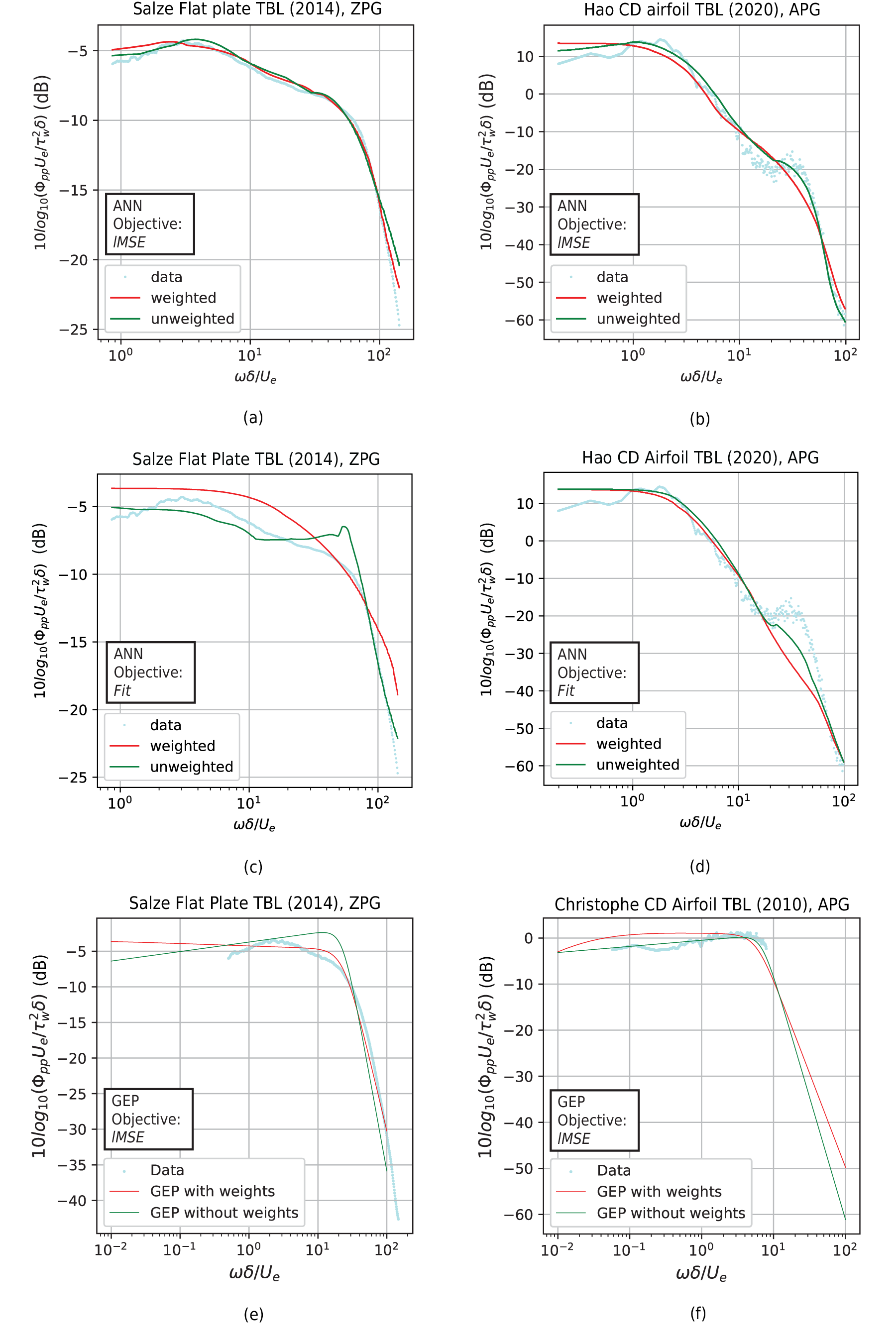}
\caption{\label{fig:ann_with_without_weights_lmse_fit_trends}Salze dataset predictions suffering (a) and (c); Hao dataset predictions improving (b) and (d), from unweighted ANN training. Salze dataset predictions suffering (e), and Christophe dataset predictions improving (f) from unweighted GEP training with $lMSE$ as the objective function}
\centering
\end{figure*}

The effect of different objective functions ($lMSE$ and $Fit$) on the predictions of wall pressure spectra ($\Phi_{pp}$) using ANN is illustrated in Fig.~\ref{fig:ann_with_without_weights_lmse_fit_pred_low} (a-d). It is observed that the predictions of ANN models trained with weighted/unweighted $lMSE$ as the objective function are more accurate than the models trained with weighted/unweighted $Fit$. Hence, we use $lMSE$ as the objective function for the rest of the study unless otherwise mentioned. Since GEP predicts a different mathematical expression every run, 10 trials with 100 individuals each, having a head length = 4, and optimization probability 0.05 were evolved over 5000 generations with weighted/unweighted $lMSE$ as the objective function. Figures ~\ref{fig:ann_with_without_weights_lmse_fit_pred_low} (e) and (f) plot the Predicted vs True values of PSD ($\Phi_{pp}$) obtained from the corresponding best-performing expressions which are listed below:

Weighted:

\begin{equation}\label{eq:4}
    \widetilde{\Phi}_{pp} = \dfrac{0.46\widetilde{\omega}^{0.99}(\beta+1)^{1.5}}{\dfrac{R_T^{0.55}\widetilde{\omega}^{1.02}}{\Pi^{2.76}+\Delta}+M\Delta^{6.75}\left(\dfrac{R_T+\widetilde{\omega}^{5.07}}{R_T^{5.84}}\right)}
\end{equation}

and Unweighted:

\begin{equation}\label{eq:5}
    \widetilde{\Phi}_{pp} = \dfrac{\widetilde{\omega}\left(\Pi^{6.71}+(\beta+1)^{1.4}\right)}{\widetilde{\omega}^{0.87}\Delta^{0.98}M^{0.44}(1+M^{0.97})+\dfrac{\widetilde{\omega}^{5.31}}{R_T^{7.07}}(\Pi^{5.31}+\widetilde{\omega}\Delta^{6.09})}
\end{equation}

Interestingly, Fig. ~\ref{fig:ann_with_without_weights_lmse_fit_pred_low} shows that ANN and GEP models trained with unweighted objective functions resulted in a lower $lMSE$ error than those with the weighted ones. Although this is an unexpected trend, Table ~\ref{tab:table51} delves into the performance of the models with weighted objective functions on the minority datasets. As intended, the prediction of datasets in the minority experiment (Salze) has indeed improved with weighted objective functions. In contrast, the accuracy of the datasets from a majority experiment (Christophe) has suffered due to the choice of weights. Figure \ref{fig:ann_with_without_weights_lmse_fit_trends} plots the wall pressure spectra of the majority and minority datasets to further demonstrate the effect of weighted/unweighted objective functions across different training schemes. The results show that the weighted objective functions significantly improve predictions of minority datasets (Salze). This trend is particularly observed for poorly converging training methods such as ANN with $Fit$ or the GEP. While the prediction accuracy of the dominant datasets (Christophe and Hao) decreases with the weighted training schemes, the spectra are not far off. On the other hand, unweighted schemes are observed to overfit the dominant datasets retaining their features and deteriorating the accuracy of the minority dataset predictions (refer to the predictions of inertial subrange in Fig \ref{fig:ann_with_without_weights_lmse_fit_trends}(c) and (d)). These effects, however, become less pronounced in the case of an aggressively converging training scheme such as ANN with $lMSE$ where both the weighted/unweighted datasets result in competitive predictions.

The aforementioned discussion highlights the importance of optimizing the weights assigned to skewed datasets which will (a) minimize any bias in the predictions of ML models and (b) improve the performance of poorly converging models as it allows the training algorithm to focus more on minority datasets.

\begin{figure*}
\centering
\includegraphics[scale  = 0.8]{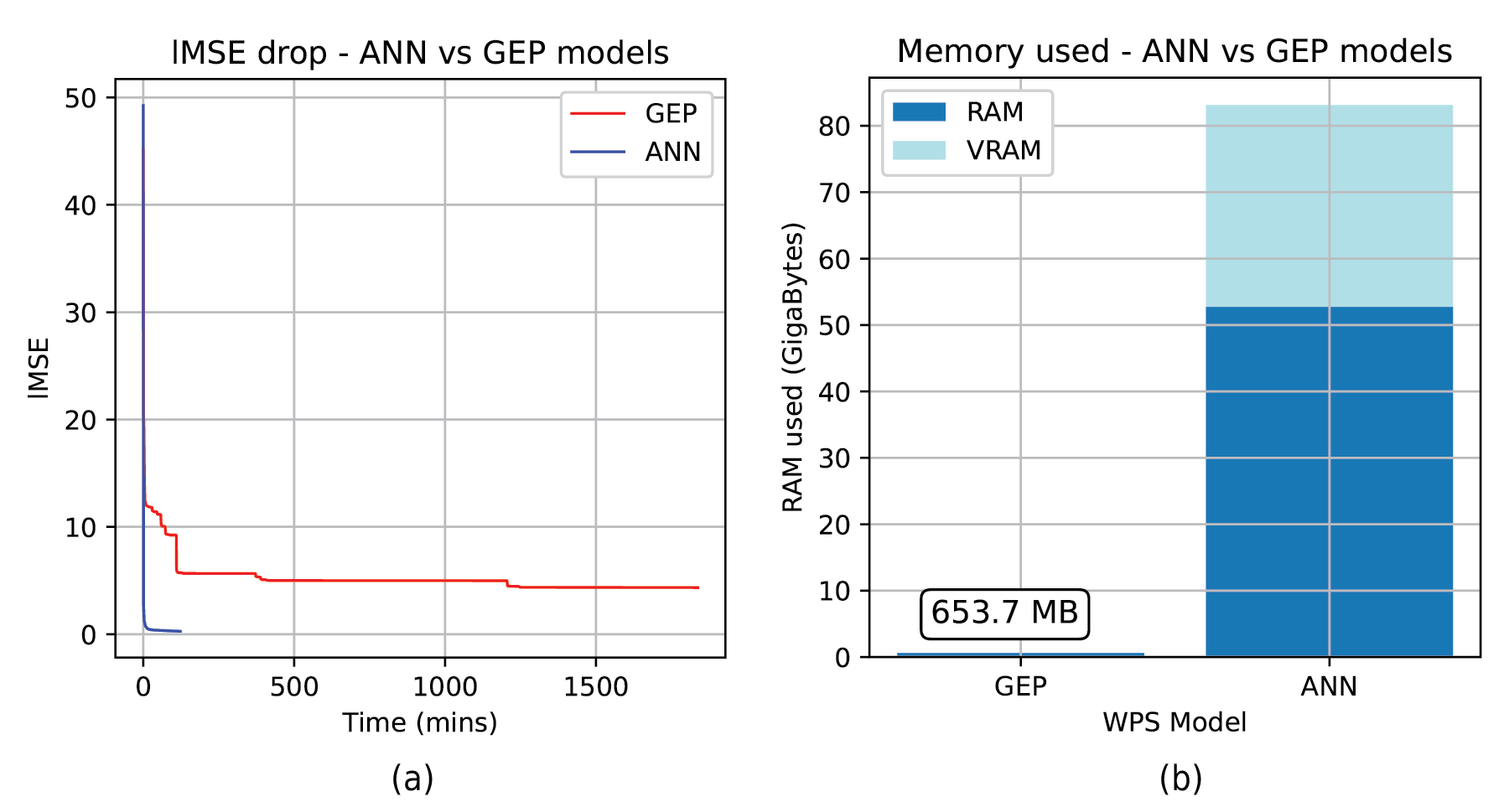}
\caption{\label{fig:time_space_complexity}(a) Model performance evolution through time, and  (b) the RAM and VRAM used by the ANN and GEP models while training}
\centering
\end{figure*}

\subsection{\label{ch3:sec2}ANN vs GEP}

The present section compares the computational efficacy of GEP and ANN in predicting the wall pressure spectra. Both the algorithms are trained against the same input data and with the same objective function (weighted $lMSE$, Eq. ~\ref{eq:-1}). The trials are conducted on a machine equipped with a dual node Intel(R) Xeon(R) Gold 5120 CPU @ 2.20GHz, with 14 cores, and an Nvidia Tesla V100 32GB GPU.

Figure ~\ref{fig:time_space_complexity} (a) compares the convergence of ANN and GEP models with time. It is evident that the ANN achieves superior results in a shorter duration. This is in contrast to GEP which struggles to converge to the $lMSE$ achieved by ANN even after a significantly longer training. With GEP, the best-fitting model from a pool may or may not improve over the generations thereby resulting in a `stepwise' drop in $lMSE$. The mechanism for evolving a best-fitting model in GEP is mutation and the other genetic operators, but they do not necessarily result in a better model and are therefore considered diverging in nature. As a result, the next best-fitting model can appear anytime through evolution. While there are ways to control the divergence induced by mutation, such as balancing it out with aggressive selection or using a local optimization function, in general, the `stepped' convergence still follows.

Figure ~\ref{fig:time_space_complexity} (b) compares the virtual memory consumption with the present implementation of both algorithms. Space consumption of GEP is dependent on the size of the population, the length of each chromosome (which are basically fixed-length strings) and, additional storage required for performing actions such as evaluation, selection and, mutation. All these operations are inexpensive when compared to the ANN training. Although the memory occupied by the trainable parameters in ANN is minimal, the training process itself involves supervised learning that requires storage of gradients, and other intermediate computations that are memory intensive. Parallel computing using GPUs does help out in reducing this requirement by allowing for larger batch sizes that reduce the required number of forward and backward passes.

\begin{figure*}
\centering
\includegraphics[width=\linewidth]{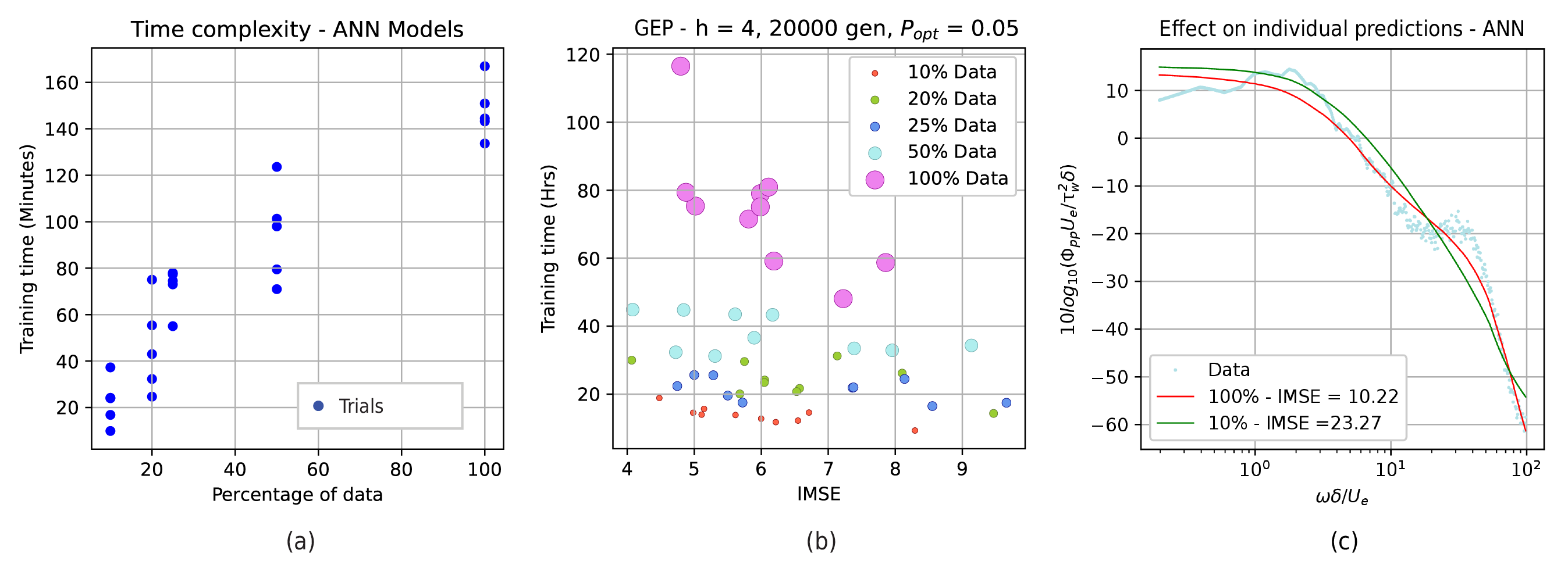}
\caption{\label{fig:ann_gep_time_space} Lower training times achieved with low-resolution training data - (a) ANN, (b) GEP. (c) Effect of low-resolution training data on ANN predictions for datasets with complex trends}
\centering
\end{figure*}

Figures ~\ref{fig:ann_gep_time_space} (a) and (b) compare the training time required with a reduced data resolution. As expected, both algorithms require less training time when trained against the same input data that has a lower resolution. With ANN, five trials were conducted to account for variation in the reported training time and one can observe a monotonic drop in training time with coarser input data. This is much easier to establish in the case of ANN models since they all converge aggressively resulting in competitive fits (see Fig.~\ref{fig:ann_gep_time_space} (c)). Performing a similar analysis with GEP involves some practical complications. Different trials of the GEP algorithm require different runtimes to arrive at models with similar performance. It highlights the need to stop GEP training after a predefined number of generations although the models obtained this way can have different performances. Figure ~\ref{fig:ann_gep_time_space} (b) reports the $lMSE$ values from the 10 trials for each resolution of the sampled data. The plot shows that cases with coarser data input, in general, require less time for training and can still result in competitive predictions. To summarize, it seems beneficial to train models (specifically GEP) with coarsely sampled data, although ensuring that the local features of the data are not lost in the re-sampling process. 


\section{\label{ch4}Guided search in Gene Expression Programming}

GEP and ANN algorithms update the model parameters through every epoch to accurately fit the data. The superior performance of ANN (in terms of speed and reliability) is attributed to training algorithms like stochastic gradient descent that guide the direction in which the ANN model parameters should be updated. On the other hand, GEP updates the models through genetic operators which rely on the evolutionary approach of trial and error. Although this approach allows for a diverse search space, it is computationally expensive. Oftentimes, with some of the known dependencies, the search space can be reduced resulting in faster convergence. This guided search approach also facilitates the discovery of complex trends in the rest of the search space. The following subsections demonstrate several guided search strategies that have been explored to accelerate GEP training and reliability.

\subsection{\label{ch4:sec2}GEP Training Schemes}

In contrast to the baseline scheme that uses the raw data, the current section introduces three additional schemes: Noise reduction with ANN filter, Omega2, and Gene4, to accelerate the GEP training.  

\subsubsection*{\label{ch4:sec2:subsec1}Raw data scheme}
This is a baseline GEP training scheme that uses the default noisy raw data and the linking function, Eq. \ref{eq:1002}, to train the GEP population.

\subsubsection*{\label{ch4:sec2:subsec2}Noise reduction with ANN filter}

Noisy input datasets can misguide GEP towards sub-optimal solutions where the resulting model fails to recognize the underlying trend. The absence of feedback in mutations increases the training time required for the GEP model to converge to a global optimum. Smoothing techniques, such as kernel smoothing or moving averages, can be used to reduce the noise although these might suffer from issues like uneven noise filtering and might filter important uncertainties in parts of the data. On the other hand, Artificial neural networks are well known for their ability to handle the trade-offs between retaining information and filtering noise. Sections \ref{ch3:sec1} and \ref{ch3:sec2} have also demonstrated their ability to produce well-performing models of WPS with reasonable training times. Hence, we explore the strategy of ANN-filtered input data to assist the GEP algorithm. 


\subsubsection*{\label{ch4:sec2:subsec3}Omega2 Scheme}

The canonical shape of the WPS remains more or less the same across TBLs and hence a good guess of the frequency dependence of PSD in the respective regimes can be made. For example, one of the popular WPS models by Goody has three parts A, B, and C:

\begin{equation}\label{eq:1001}
\widetilde{\Phi}_{pp}= \dfrac{C_{2} \widetilde{\omega}^{2}}{\left(\widetilde{\omega}^{0.75}+C_{1}\right)^{3.7}+\left(C_{3} \widetilde{\omega}\right)^{7}} = \dfrac{A}{B+C}
\end{equation}

Where,

$$
C_{1} = 0.5;\; C_{2} = 3.0;\; C_{3} = 1.1R_T^{-0.57}
$$

The numerator $A$ dominates at low frequencies ($\widetilde{\omega} \rightarrow 0$), resulting in the $\widetilde{\omega}^2$ trend, which is particularly valid under zero pressure gradients (ZPG) \cite{Goody2004}. The exponent of $\widetilde{\omega}$, however, changes with pressure gradients as observed in Figures \ref{fig:vki_data_rescaled} and \ref{fig:wps_trends}. For example, WPS of a TBL subjected to APG exhibits a $\approx \widetilde{\omega}^{0.5}$ rise at low frequencies (see Fig. \ref{fig:wps_trends} from experiments of Salze \emph{et al.}\cite{Salze2014}). While both $B$ and $C$ in Eq. \ref{eq:1001} dictate the mid-frequencies, the term $C$ governs trends at higher frequencies. Goody's model exhibits a $\widetilde{\omega}^{-5}$ trend at higher frequencies which is consistent with those observed in Fig. \ref{fig:wps_trends}. Unlike the other contemporary models in literature \cite{fritsch_2023}, Goody's model is stiff and reasonably retains its shape over a range of flow conditions.

As discussed above, the shortcomings of Goody's model which only depends on $R_T$ (in addition to $\widetilde{\omega}$) are apparent under pressure gradients and non-canonical flows. GEP algorithm is effective in seeking additional contributing variables and accelerates the development of semi-empirical models like Goody. To assist GEP search with the inherent non-linearity of the WPS problem, Dominique \emph{et al.}\cite{Dominique2021} prescribed the following trigenic linking function, Eq. \ref{eq:1002} which is analogous to Goody's model:

\begin{equation}\label{eq:1002}
    Y^{GEP} =  \dfrac{\mathit{Sub-ET}_1}{\mathit{Sub-ET}_2+\mathit{Sub-ET}_3}
\end{equation}
Where $\mathit{Sub-ET}_i$ represents a sub-expression interpreted from a gene in the chromosome and $Y^{GEP}$ represents the output model.

GEP still has a tough task of working out the correct combination, especially the strong dependence on frequency ($\widetilde{\omega}$) across different regimes. Since the WPS at low frequencies follows a $\widetilde{\omega}^2$ trend for ZPG flows, we explore the benefit of premultiplying Eq. \ref{eq:1002} with $\widetilde{\omega}^2$ to build that dependency right into the solution. The scheme further uses assistance from ANN-filtered data (modified inputs) for improved predictions. Hereafter, this strategy will be referred to as the $Omega2$ $scheme$ which is implemented by modifying the objective function as follows: 

\begin{equation}\label{eq:1003}
    \mathit{lMSE} =  \dfrac{1}{N}\sum\limits_{i=1}^N W_i \left(10log_{10} \left(\widetilde{\omega}_i^2Y_i^{GEP}\right) - 10log_{10} Y_i^{true}\right)^2
\end{equation}

The proposed modification facilitates the algorithm to focus on finding a better combination $Sub-ET_1$ using any of the variables ($\widetilde{\omega},\Delta,H,M,\Pi,C_f,R_T,\beta$ ). Table \ref{tab:table1000} lists some of the sample models obtained using the Omega2 scheme. Interestingly, for the datasets where the low-frequency trend deviates from the $\widetilde{\omega}^2$ dependence, GEP is observed to predict a $\mathit{Sub-ET_1}$ with $\widetilde{\omega}^i$ (i being some index) to accurately fit the low frequencies.

\begin{table*}
    \centering
    \caption{\label{tab:table1000}Sample WPS models generated using omega2 scheme}
    {\renewcommand{\arraystretch}{4}
    \begin{ruledtabular}

    \begin{tabular}{ccrrr}
    Model   &$\widetilde\omega^2 \cdot Y^{GEP}$  &$lMSE$  &$\widetilde{\omega} \rightarrow 0$   &$\widetilde{\omega} \rightarrow \infty$ \\ \hline
    
    model1  &$ \widetilde\omega^2\cdot\dfrac{\widetilde{\omega}^{-1.86}\beta^{1.34}M^{-0.55}\Delta^{-0.78}\left(R_T+1\right)}{\widetilde{\omega}^{1.77}+R_TH^{1.86}+\widetilde{\omega}^{6.60}\left( \dfrac{\Delta}{R_T}\right)^{6.18}}$  &10.61   &0.14  &-6.46 \\ 
    
    model2  &$ \widetilde\omega^2\cdot\dfrac{\widetilde{\omega}^{-2}\beta^{1.42}\left(\Pi^{3.57}\widetilde{\omega}^{0.58}+R_T^{1.02}\right)}{\Delta\left(\widetilde{\omega}^{5.53} \dfrac{\Delta^{5.59}}{R_T^{6.47}}+R_TM^{0.37}\right)}$ &11.17  &0.58 &-4.95\\

    model3  &$ \widetilde\omega^2\cdot\dfrac{\widetilde{\omega}^{-1.7}\beta^{1.72}M^{-0.72}}{\widetilde{\omega}^{1.99}+\Delta\left( \dfrac{\widetilde{\omega}^{4.98}H^{8.81}}{R_T^{4.28}+\beta^{5.63}}+1\right)}$  &13.67  &0.3    &-4.68
    \end{tabular}
    \end{ruledtabular}
    }
    
\end{table*}

\subsubsection*{\label{ch4:sec2:subsec4}Gene4 Scheme}

Extending the omega2 approach, the $Gene4$ $scheme$ is aimed at further incorporating the trends at high-frequencies into the GEP models. This is achieved by modifying the original linking function in Eq. \ref{eq:1002} with (a) an $\widetilde{\omega}^2$ term in the numerator (to predict lower frequencies), (b) a $\widetilde{\omega}^7$ in the denominator (to predict higher frequencies) and (c) a timescale ratio dependency $R_T^4$ from Goody's model, a dependency also observed in other contemporary empirical models, to further assist the search. The gene4 linking function is realized using four genes as formulated below:

\begin{equation}\label{eq:1004}
    Y^{GEP}_{gene4} =  \dfrac{\widetilde{\omega}^2\mathit{Sub-ET_1}+\mathit{Sub-ET_2}}{ \dfrac{\widetilde{\omega}^7}{R_T^4}\mathit{Sub-ET_3}+\mathit{Sub-ET_4}}
\end{equation}

We have used two $Sub-ET$s in the numerator with only $Sub-ET_1$ being pre-multiplied with $\widetilde{\omega}^2$. $Sub-ET_2$ allows for searching the terms that are decoupled from the $\widetilde{\omega}^2$ dependence and facilitates a faster convergence than the omega2 scheme (as the algorithm does not have to look for a complex division to eliminate the effect of pre-multiplication). Table \ref{tab:table1003} lists some of the competitive models predicted using the gene4 scheme. It is interesting to note that the $Sub-ET_2$ term in models 1 and 2 is independent of $\widetilde{\omega}^2$ while that in model 3 is closer to $\widetilde{\omega}^2$.

\begin{table*}
    \centering
    \caption{\label{tab:table1003}Sample WPS models generated using gene4 scheme}
    {\renewcommand{\arraystretch}{4}
    \begin{ruledtabular}

    \begin{tabular}{ccrrr}
    Model   &$ \dfrac{\widetilde{\omega}^2\mathit{Sub-ET_1}+\mathit{Sub-ET_2}}{ \dfrac{\widetilde{\omega}^7}{R_T^4}\mathit{Sub-ET_3}+\mathit{Sub-ET_4}}$  &$lMSE$  &$\widetilde{\omega} \rightarrow 0$   &$\widetilde{\omega} \rightarrow \infty$ \\ \hline
    
    model1  &$ \dfrac{\widetilde{\omega}^2\left( \dfrac{\beta^{2.45}\Pi^{5.13}}{R_T^{1.22}}+ \dfrac{0.015}{M}\right) +\left( \dfrac{C_f}{M^{2.08}}+\beta^{2.84}\right)\left( \dfrac{R_T}{M^{0.14}}\right)}{ \dfrac{\widetilde{\omega}^7}{R_T^4}\left( \dfrac{\Delta^{2.89}H^{7.62}}{R_T\omega^{1.12}} \right)+\left(R_T^{1.29}+\Delta^{2.03}\right)\left(M^{0.34}+\beta\right)}$  &10.41   &2  &-3.88 \\ 
    
    model2  &$ \dfrac{\widetilde{\omega}^2\left( \dfrac{\Pi^{9.74}+\widetilde{\omega}^{0.65}}{\Delta^{5.02}}\right)\left(0.1R_T^{1.27}\right)+\beta\left(M^{-1}\Delta^{-1}+\beta^{1.21}\right)}{ \dfrac{\widetilde{\omega}^7}{R_T^4}14.2M^{2.96}\left(M^{1.24}+1\right)+\widetilde\omega+5.2\widetilde\omega^{-0.35}}$ &10.95  &3 &-4.35\\

    model3  &$\dfrac{\widetilde{\omega}^2\left(\dfrac{\beta^{2.43}\left(\Delta+\Pi\right)}{H\Delta^{1.15}} \right)+\dfrac{\omega^{2.64}}{M\left(R_T+\Pi^{0.06}\right)}}{\dfrac{\widetilde{\omega}^7}{R_T^4}\left(\dfrac{M\Delta^{6.79}}{R_T^{2.39}} \right)+\widetilde\omega^{1.55}\left(H+\widetilde\omega^{0.94}\right)}$  &12.23  &2.64    &-4.36
    \end{tabular}

    \end{ruledtabular}
    
    }

\end{table*}

\subsection{\label{ch4:sec3}Comparisons}

In this section, we compare the performance of the aforementioned schemes in terms of (a) convergence rate during training and (b) model accuracy at the end of GEP evolution. For each of the four schemes, 10 trials were conducted on a pool of 100 individuals, with a head length ($h$) of 4 trained over 20000 generations. The input data is resampled to 25\% of the original resolution and for every iteration, the population passed through the least squares optimization loop with an optimization probability of $P_{opt} = 0.05$. We probe into the temporal evolution of the training statistics with weighted $lMSE$ as the objective function.

\begin{table}
\centering
    \caption{\label{tab:table1005}Nomenclature for y-axis quantities in Fig. \ref{fig:ensembled_lmse_fit_comparison} (a-e)}
\begin{ruledtabular}

\begin{tabular}{ll}

\textbf{Quantity}    & \textbf{Description}\\ \hline
$<y>_E$     & Ensemble average of 10 trials\\ \hline
$median(y)$ & Median of 10 trials\\ \hline
$y_{25-75}$ & 25-75 percentile band of 10 trials\\ \hline
$y_{max}$   & Maximum value of 10 trials\\ \hline
$y_{min}$   & Minimum value of 10 trials\\ 
\end{tabular}
\end{ruledtabular}
\end{table}

\begin{figure*}
\centering
\includegraphics[scale = 1]{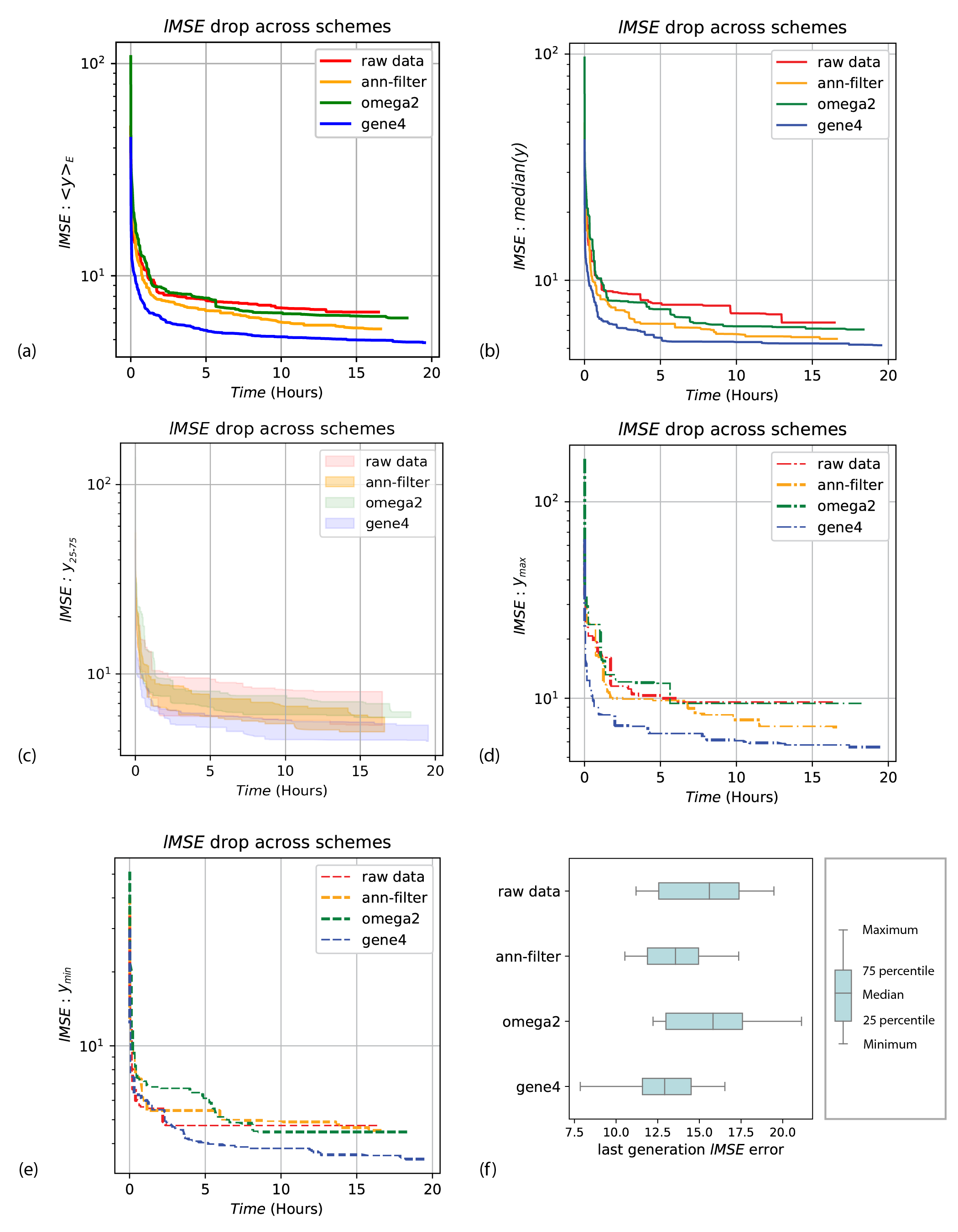}
\caption{\label{fig:ensembled_lmse_fit_comparison} Comparison of fitness distribution of the best individuals across 10 trials with $lMSE$ as the objective function for different schemes. Evolution of following statistics: (a) $<y>_E$, (b) $median(y)$, (c) $y_{25-75}$ (d) $y_{max}$ (e) $y_{min}$, of GEP trials are plotted against time. (f) Box plot comparing unweighted $lMSE$ for the last generation, showing the scatter in $lMSE$ from minimum to maximum across the trials. The blue box represents the trials with $lMSE$ between the $25-75$ percentile and the intermediate line shows the median value.)}
\centering
\end{figure*}

\begin{table*}
\centering
    \caption{\label{tab:table10001}Quantitative overview of the evolution of the relative performance of the proposed GEP training schemes with respect to the baseline raw data scheme. This relative performance is evaluated every five hours into the training phase.}
\renewcommand{\arraystretch}{1.2}
\begin{ruledtabular}   

\begin{tabular}{l|rrrrrrrrrrrrrrr}
\multirow{3}{*}{Scheme} & \multicolumn{15}{c}{Percentage $lMSE$ change w.r.t raw scheme}                                                                                                                                                                                                                                                                                                                                                                                  \\ \cline{2-16} 
                        & \multicolumn{3}{c|}{$<y>_E$}                                                         & \multicolumn{3}{c|}{$median(y)$}                                                     & \multicolumn{3}{c|}{$y_{25-75}$}                                                     & \multicolumn{3}{c|}{$y_{max}$}                                                       & \multicolumn{3}{c}{$y_{min}$}                                                       \\ \cline{2-16} 
                        & \multicolumn{1}{l}{5 hrs} & \multicolumn{1}{l}{10 hrs} & \multicolumn{1}{l|}{15 hrs} & \multicolumn{1}{l}{5 hrs} & \multicolumn{1}{l}{10 hrs} & \multicolumn{1}{l|}{15 hrs} & \multicolumn{1}{l}{5 hrs} & \multicolumn{1}{l}{10 hrs} & \multicolumn{1}{l|}{15 hrs} & \multicolumn{1}{l}{5 hrs} & \multicolumn{1}{l}{10 hrs} & \multicolumn{1}{l|}{15 hrs} & \multicolumn{1}{l}{5 hrs} & \multicolumn{1}{l}{10 hrs} & \multicolumn{1}{l}{15 hrs} \\ \hline
ann-filter              & -10.4                     & -14.4                      & \multicolumn{1}{r|}{-15.8}  & -18.5                     & -19.0                      & \multicolumn{1}{r|}{-13.8}  & -57.6                     & -49.0                      & \multicolumn{1}{r|}{-58.6}  & -2.7                      & -18.8                      & \multicolumn{1}{r|}{-24.4}  & 5.4                       & 4.9                        & 4.6                        \\
omega2                  & 2.7                       & -5.3                       & \multicolumn{1}{r|}{-4.1}   & -5.9                      & -12.3                      & \multicolumn{1}{r|}{-5.9}   & -57.7                     & -37.7                      & \multicolumn{1}{r|}{-54.0}  & 18.7                      & -1.8                       & \multicolumn{1}{r|}{-1.8}   & 29.0                      & -5.9                       & -5.9                       \\
gene4                   & -27.6                     & -27.0                      & \multicolumn{1}{r|}{-26.1}  & -29.5                     & -25.3                      & \multicolumn{1}{r|}{-19.2}  & -76.0                     & -52.2                      & \multicolumn{1}{r|}{-59.4}  & -33.9                     & -36.1                      & \multicolumn{1}{r|}{-39.4}  & -14.9                     & -19.7                      & -24.4                     
\end{tabular}
\end{ruledtabular}
\end{table*}

For different training schemes, Fig. \ref{fig:ensembled_lmse_fit_comparison} (a-e) compares the temporal evolution of the resampled training statistics (see Appendix \ref{appendix}) from the 10 trials using $lMSE$ as the objective function. These metrics include $<y>_E$, $median(y)$, $y_{25-75}$, $y_{max}$, and $y_{min}$ which are briefly described in Table \ref{tab:table1005}. For every trial, the best-performing model at the end of the training is recorded. Figure \ref{fig:ensembled_lmse_fit_comparison} (f) compares the $lMSE$ distributions of these best models across different schemes. Table \ref{tab:table10001} provides a quantitative overview of the relative performance of these schemes with respect to the baseline scheme, probed every 5 hours into the training period. The following inferences can be drawn from the statistics presented in Fig. ~\ref{fig:ensembled_lmse_fit_comparison} and Table \ref{tab:table10001}:

\begin{itemize}
    \item The gene4 scheme predicts more accurate models than other schemes throughout the evolution. When compared to the baseline scheme, Table \ref{tab:table10001} shows the superior performance of gene4 with a consistent drop of $\approx 25\%$ in its ensemble averaged $lMSE$ and $\approx 20\%$ drop in its median $lMSE$. A $\geq \approx 60\%$ drop of $lMSE$ within the interquartile range, $y_{25-75}$ with the gene4 scheme suggests that the scheme predicts more competitive models at any instance throughout the evolution. Compared to the baseline scheme, the $y_{max}$ $lMSE$ (worst-performing model) and $y_{min}$ $lMSE$ (best-performing model) from the gene4 scheme trials exhibit a $\ge 30\%$ and $\ge 15\%$ lower value respectively. Trends in figure \ref{fig:ensembled_lmse_fit_comparison}(a-e) also illustrate that the aforementioned improvements from the gene4 scheme are consistently observed throughout the evolution which implies that the gene4 scheme converges faster than the rest. Since the evolution of different training schemes can stop at different times, Figure \ref{fig:ensembled_lmse_fit_comparison}(f) compares the $lMSE$ statistics of the last generation. Across trials, gene4 scheme has a best-performing model with $\approx 30\%$ drop in the $y_{min}$ $lMSE$, a $\approx 17\%$ drop in the median $lMSE$ and a $\approx 40\%$ reduction in $lMSE$ within the interquartile range, $y_{25-75}$ implying that gene4 scheme predicts more accurate models with improved reliability.
    \item Statistical improvements in $lMSE$ obtained using the ann-filter scheme are similar to those of the gene4 scheme albeit to a lesser extent.
    \item In contrast, the omega2 scheme yields either marginal or inconsistent gains over the raw data scheme throughout the training resulting in arguably worse-performing last-generation models (see Figure \ref{fig:ensembled_lmse_fit_comparison}(f)). This indicates that multiplying the entire numerator ($Sub - ET_1$ of Eq. \ref{eq:1002}) with $\widetilde\omega^2$ has resulted in poorly optimized expressions. As seen in the case of model 2 in Table. \ref{tab:table1000}, the algorithm has probably worked more towards decoupling the frequency terms.
\end{itemize}

Considering all the inferences, it can be concluded that the strategies involved in the development of the gene4 scheme (which include using ann-filtered input data, built-in low and high-frequency trends, and a modified linking function) are justified. Hence, the Gene4 scheme serves as a faster, and more accurate alternative to the baseline scheme while inferring models out of WPS data using the GEP algorithm.

\subsection{\label{ch4:sec4}Validation}

\begin{figure*}
\includegraphics[scale = 0.85]{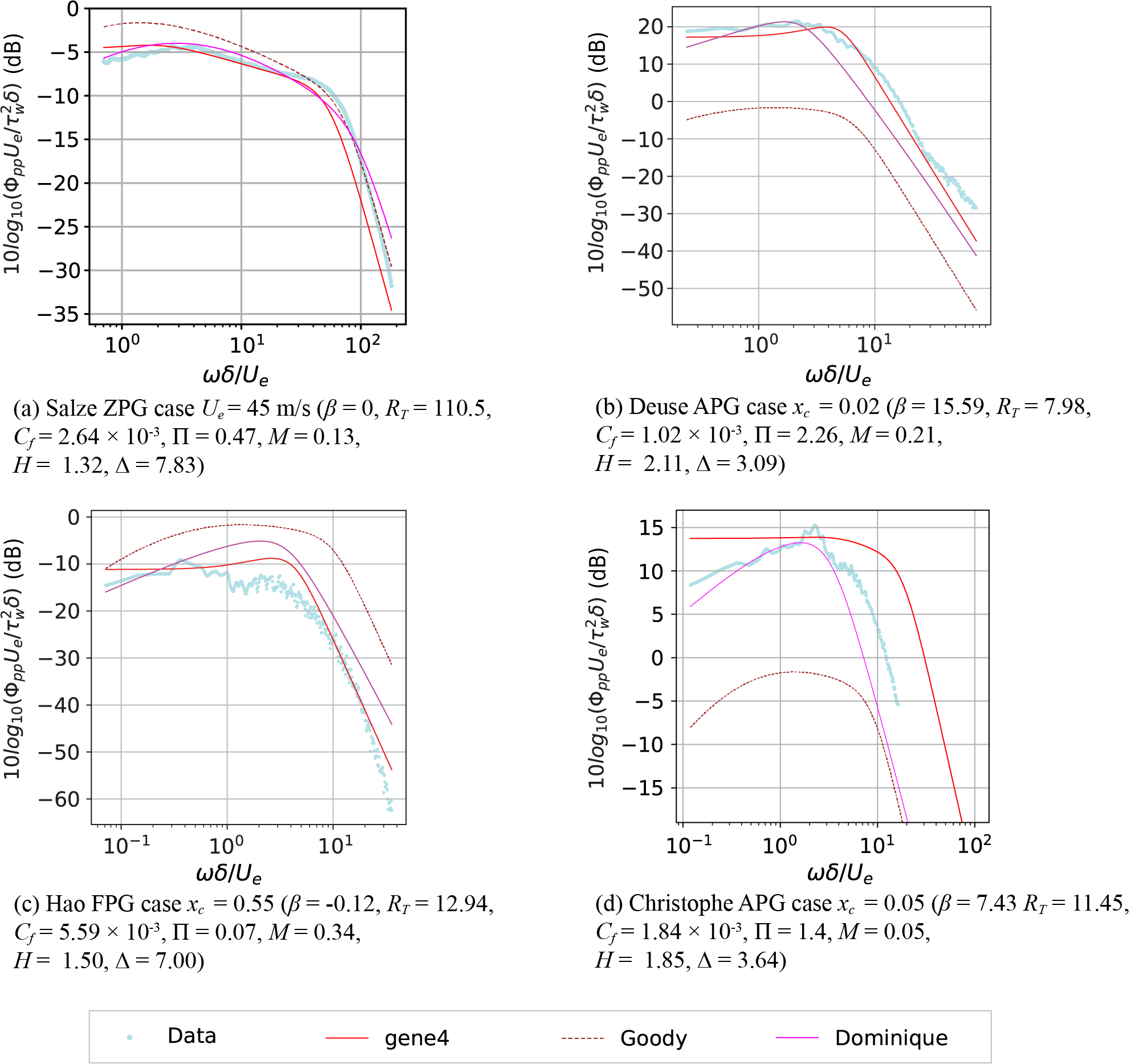}
\caption{\label{fig:validation}Performance a sample GEP model trained with gene4 scheme on unseen datasets.}
\centering
\end{figure*}

In this section, we evaluate the performance of the gene4 scheme on unseen data to ensure that the scheme predicts the mathematical expressions of appropriate complexity. Four datasets are discarded from each experiment from the complete database. Gene4 scheme was trained on the rest of the data with the same hyperparameter setup used in the previous section. One of the models within the interquartile range, $y_{25-75}$, is presented below:

\begin{equation}\label{eq:1006}
    \widetilde\Phi{pp} = \dfrac{\widetilde\omega^2\left(\dfrac{\Pi^{8.05}}{\Delta}+0.22\beta^{1.54}\right)+0.69\beta^{2.66}+\Pi^{8.22}}{\dfrac{\widetilde\omega^7}{R_T^4}(M^{4.07}(100(10+\Delta)))+(M^{0.44}(H^{5.81}+\widetilde\omega^{2.37}))}
\end{equation}

Equation \ref{eq:1006} shows that the model retains the built-in frequency trends and incorporates additional complexities of the pressure gradient ($\beta$) and Mach number dependence. Figure \ref{fig:validation} plots the corresponding predictions of the unseen datasets. It is evident from Figure \ref{fig:validation}(a-c) that the model presented surpasses the predictions of both Goody's model and Dominique's GEP model (which has been trained over the entire database). This ensures that the model does not overfit the data, which is in fact an encouraging step towards the development of a generalized WPS model that can reasonably predict unseen data.
However, Fig. \ref{fig:validation} (d) illustrates that, although the model performance is superior to that of Goody, there is scope to improve it further. For example, the appearance of coefficients such as 100 and 10 in this sample model shows that at times the GEP algorithm may struggle to find correct coefficients from the RNC array (Section \ref{ch2:sec4:subsec1:subsubsection:8}) resulting in a sub-optimal mathematical solution. However, the same gene/$Sub - ET_3$ does not hold a frequency term implying that the built-in trend by gene4 is valid, and with more mutations, GEP can find a superior alternative. 

\subsection{\label{ch4:sec5}Stepped schemes}
\begin{figure}
\centering
\includegraphics[scale = 0.6]{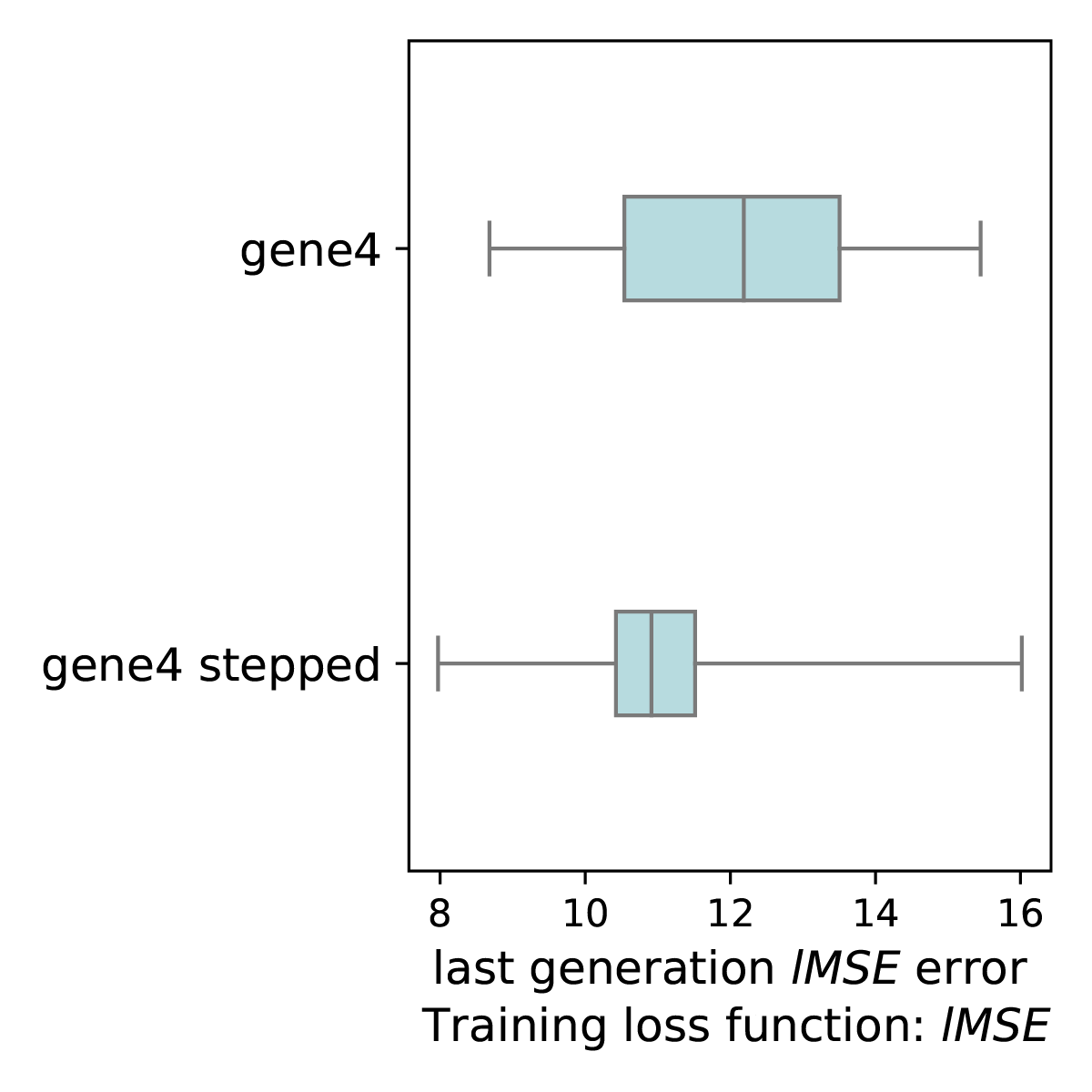}
\caption{\label{fig:last_iteration_2stp_lmse_40k}
Box plot comparing unweighted $lMSE$ for the last generation ($40000^{th}$), showing the scatter in $lMSE$ from minimum to maximum across the trials. The blue box represents the trials with $lMSE$ between the 25 - 75 percentile and the intermediate line shows the median value.}
\centering
\end{figure}

\begin{figure*}
\centering
\includegraphics[scale = 1]{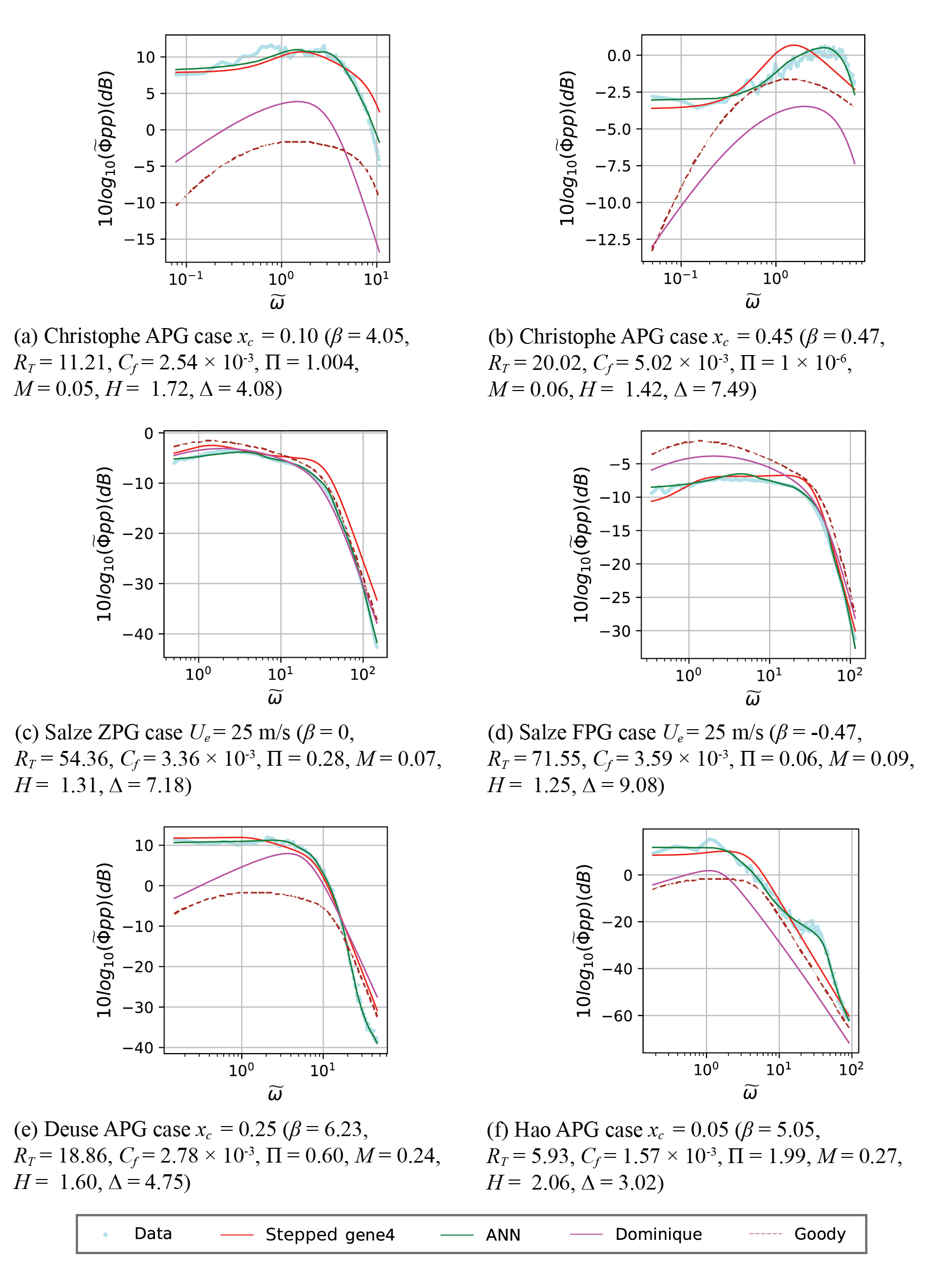}
\caption{\label{fig:complex_results} Dataset-wise predictions of a GEP model, Eq. \ref{eq:1007}, trained with the stepped training scheme and comparison with ANN predictions along with the predictions of Dominique's GEP model and Goody's model}
\centering
\end{figure*}

\begin{figure*}
\centering
\includegraphics[scale = 0.9]{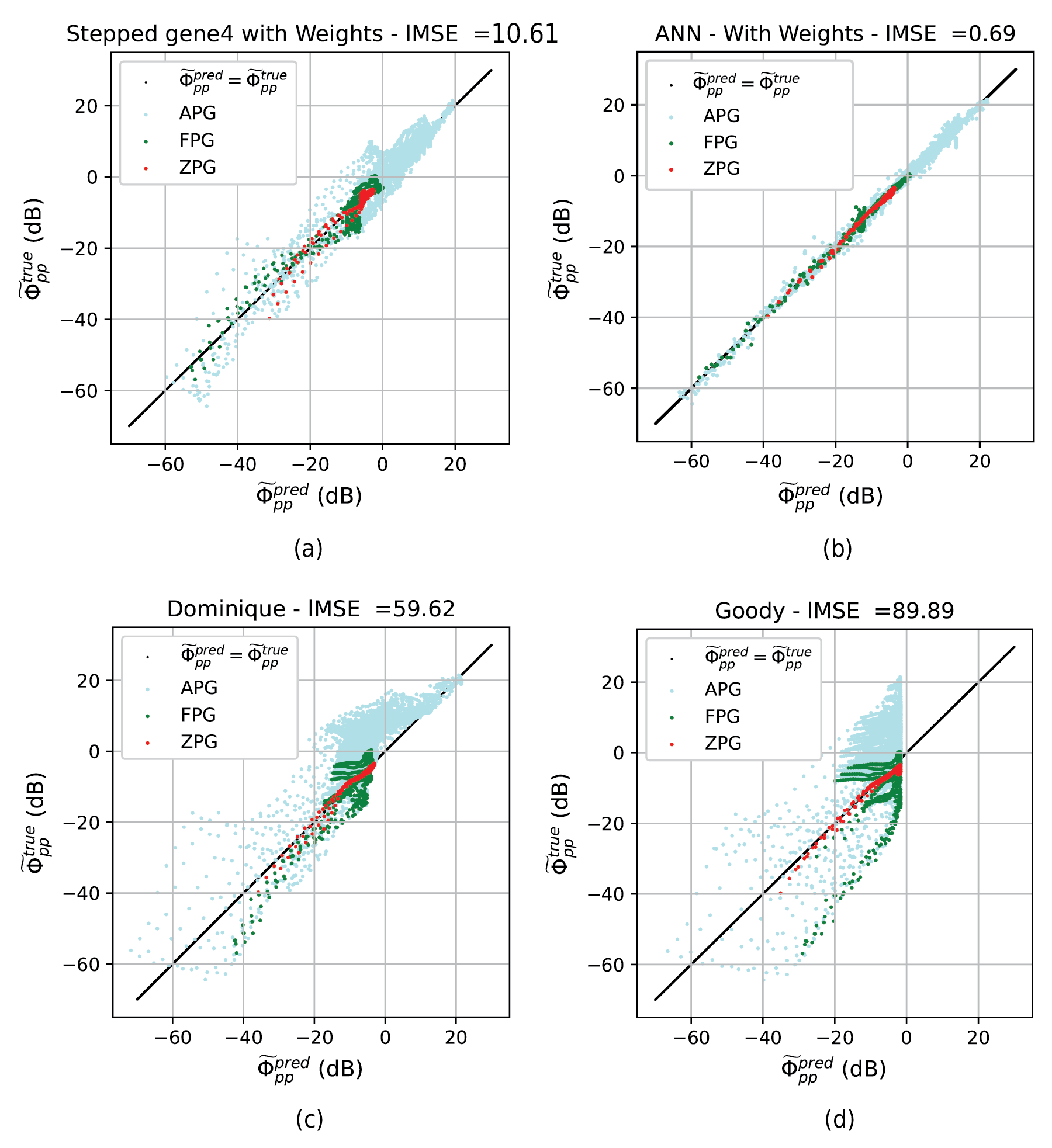}
\caption{\label{fig:stepped_ann_dominique_goody}Predictions of a GEP model, Eq. \ref{eq:1007}, trained with the stepped training scheme and comparison with ANN predictions along with the predictions of Dominique's GEP model and Goody's model. (Every tenth point is reported for plotting purposes)}
\centering
\end{figure*}

We have further explored the strategy of stepped training schemes which is analogous to the idea of learning a language. One should learn the basic features of the language first, to understand the intricate constructions in the literature. Likewise learning the basic trends of the datasets that show a close resemblance to the canonical WPS is a priority. Once the GEP population learns to fit these priority datasets satisfactorily it is allowed to further improve the formulations by exposing it to a dataset that has similar canonical features with added complexities. In the present work, we train the GEP algorithm in two steps. In step 1, we prioritize exposing the population to Salze datasets which have the highest similarity to the canonical form of WPS. After the first step, the entire population, which also includes the best individual, was saved. In step 2, the entire population saved in step 1 is exposed to the complete dataset for further training. For each step, the GEP algorithm was trained with the same hyperparameter environment detailed in Section \ref{ch4:sec3}. Figure \ref{fig:last_iteration_2stp_lmse_40k} shows the stepped gene4 scheme exhibiting superior accuracy and reliability over its non-stepped counterpart with $\approx 10\%$ reduction in the median (from 12.2 to 10.9) and $\approx 60\%$ drop in the $lMSE$ spread of the interquartile range (from 3.0 to 1.1). Hence, it can be deduced that at least half of the models trained using the stepped gene4 scheme are competitive with each other. A sample model, with $lMSE = 10.61$ lying within the interquartile range, reads as follows:
\begin{equation}\label{eq:1007}
    \widetilde\Phi_{pp} = \dfrac{\widetilde\omega^{3.12}+\beta^{2.14}\widetilde\omega^{2.41}+\left(\dfrac{C_f}{M^{4.42}\Delta^{2.34}}\right)\widetilde\omega^2+\beta^{2.14}R_T^{0.29}}{(M^{3.2}\Delta^{6.35}+R_T)\dfrac{\widetilde\omega^{7.71}}{R_T^5}+\Delta^{0.9}(\widetilde\omega^{2.97}+H^2)}
\end{equation}

where the low and the high frequency exponent predictions are, 

$$
\widetilde\omega \rightarrow 0 = \widetilde\omega^{3.12}; \; \widetilde\omega \rightarrow \infty = \widetilde\omega^{-4.59}.
$$


Figure \ref{fig:complex_results} further compares the WPS predictions of the sample GEP model (Eq. \ref{eq:1007}) obtained using stepped gene4 scheme across different datasets. Its superior accuracy as compared to Goody's model and Dominique's GEP model is apparent. In particular, Figures \ref{fig:complex_results} (a,b,e) illustrate the high complexity captured by this model. Interestingly, its frequency trends and magnitude are in close agreement with those predicted by the ANN model. Figure \ref{fig:complex_results} (c) assures that its predictions of the canonical WPS are unaffected and competitive with other models. Despite its superior performance, Fig. \ref{fig:complex_results} (b), (d), and (f), show that the model fails to capture some local features and under or over-predicts the trends in certain datasets.

The scatter plot in Figure \ref{fig:stepped_ann_dominique_goody} compares the predicted vs true $\widetilde\Phi_{pp}$ values for the entire dataset comprising different flow conditions (APG, FPG, and ZPG). The aforementioned GEP model trained with the stepped gene4 scheme significantly improved the predictions across the entire dataset resulting in a $lMSE$ of 10.6 in contrast to $lMSE$ of $\approx 60$ with Dominique's GEP model and $\approx 90$ with Goody's model. In particular, the scatter in the predictions of datasets with APG and FPG has considerably reduced when compared to Dominique's GEP model and Goody's model. Despite these improvements, the $lMSE$ of the proposed GEP model is still an order of magnitude higher than that of the ANN. It implies that there is further scope to develop a generalized WPS model using GEP through improvements in training strategies.

\section{\label{conclusions}Conclusions}

The study presents a machine learning-based framework that uses data-driven modeling to predict wall pressure spectra (WPS) underneath turbulent boundary layers. Different datasets of WPS from experiments and high-fidelity numerical simulations covering a wide range of pressure gradients and Reynolds numbers are considered for training. 
This dataset however appears to be skewed and a cosine-similarity matrix has been used to quantify the visual resemblance of WPS trends across the experiments. The efficacy of two machine learning techniques, namely artificial neural networks (ANN) and gene expression programming (GEP) is evaluated. Firstly, an optimal hyperparameter environment is identified that yields the most accurate predictions for the respective ML methods. This includes assessing the effect of objective functions ($LMSE/Fit$) on the convergence rate of ANN models. Of these, $Fit$ (which is a multi-objective function comprising both $lMSE$ and mean squared error) is shown to converge at a slower rate. Interestingly, it has been observed that the prediction accuracy of such weakly converging training methods can be improved by increasing the weight of minority datasets.

For a given input database, the computational resources (training time and memory consumption) of some of the best-performing ANN and GEP models are compared. In terms of accuracy, the results show the clear superiority of ANN over GEP with the logarithmic mean squared error ($lMSE$) of ANN being less than 1 while that of GEP being around $O(10)$. The corresponding training time of ANN models is also 8 times lower ($\approx 3$ $hours$) than that of the GEP ($\approx 24$ $hours$), despite a higher memory consumption. Nevertheless, the advantage of GEP lies in predicting a realizable closed-form mathematical expression. In contrast to the unconventional structure of ANN models, these expressions from GEP can provide direct physical insight.

Novel training schemes are devised to address the shortcomings of GEP. These include (a) ANN-assisted GEP to reduce the noise in the training data (b) exploiting the physical trends in the spectra observed at low and high frequencies to guide the GEP search (c) a stepped training strategy where the chromosomes are first trained on the canonical datasets followed by the datasets with complex features. The first two training strategies are shown to accelerate the convergence and yield more consistent models ($\approx 40\%$ reduction in the spread of $lMSE$ in the interquartile range) with superior accuracy ($\approx 17\%$ reduction in the median $lMSE$). Stepped training strategy further improves upon the reliability and accuracy with an additional reduction of $\approx 60\%$ and $\approx 10\%$ in the aforementioned statistics respectively. The predictions of the resulting GEP models appear to have captured the complex trends of WPS while surpassing the accuracy of both Dominique's GEP model and Goody's model.

With the inclusion of the ever-evolving data repository of WPS, the authors believe that the methods and insights from the present study will facilitate the discovery of more consistent and generalized models. The goal is to discover GEP models that result in competitive predictions to the ones produced by the ANN models. The analytical nature of GEP models will allow for the generalization of WPS beyond the trained regimes.




\section*{Acknowledgements}

The authors wish to acknowledge NVIDIA for generously awarding Quadro P6000 and A100 GPU cards under the Academic Hardware Grant Program. The authors also wish to acknowledge the VKI team: Joachim Dominique, Jan Van den Berghe, Dr. Christophe Schram, Dr. Miguel Alfonso Mendez, Dr. Julien Christophe, and Dr. Richard D. Sandberg (from the University of Melbourne) for making their source code and data publicly available. Dr. Nagabhushana Rao Vadlamani also acknowledges the financial support from Science and Engineering Research Board (SERB), India under Mathematical Research Impact Centric Support (MATRICS) scheme (MTR/2022/000807).

\section*{AUTHOR DECLARATIONS}
\subsection*{Conflict of Interest}
The authors have no conflicts to disclose.

\section*{Data Availability Statement}

The data that support the findings of this study are available from the corresponding author upon reasonable request.

\appendix\section{}\label{appendix}
Here, we provide details of the resampling procedure which maps generations in GEP evolution with wall-clock time. At the end of every generation, we probed the Weighted objective function values (of the best individual in the generation) and the wall-clock time spent for the generation. Although each of the 10 trials is trained for a fixed set of generations, the training can take different wall-clock times. Hence, the statistics are re-sampled over a common time interval using a Piece-wise Cubic Hermite Interpolating Polynomial (PCHIP) interpolator\cite{interpolation} as illustrated in Table \ref{tab:table1004}. Although this shortens the observation window to the trial with the shortest runtime to reach 20000 generations, it facilitates the comparison of statistics in wall-clock time across different trials and training schemes. This approach of choosing the shortest window is justified since the interest lies in developing methods that converge faster.


\begin{table}[H]
    
    \centering
    \caption{\label{tab:table1004}Re-sampled GEP evolution data for statistical analysis across different schemes}
    \renewcommand{\arraystretch}{1.5}

    \begin{ruledtabular}

\begin{tabular}{|l|lllllll|}

\rowcolor[HTML]{E3DFFD} 
\cellcolor[HTML]{ECF2FF}Scheme & \multicolumn{7}{l|}{\cellcolor[HTML]{E3DFFD}Resampled time} \\ \hline

\rowcolor[HTML]{E3DFFD} 
\cellcolor[HTML]{E5D1FA}Trials & \multicolumn{1}{l|}{\cellcolor[HTML]{E3DFFD}$t_1$} & \multicolumn{1}{l|}{\cellcolor[HTML]{E3DFFD}$t_2$} & \multicolumn{1}{l|}{\cellcolor[HTML]{E3DFFD}...} & \multicolumn{1}{l|}{\cellcolor[HTML]{E3DFFD}$t_i$} & \multicolumn{1}{l|}{\cellcolor[HTML]{E3DFFD}$t_{i+1}$} & \multicolumn{1}{l|}{\cellcolor[HTML]{E3DFFD}...} & $t_n$ \\ \hline 

\rowcolor[HTML]{FFF4D2} 
\cellcolor[HTML]{E5D1FA}$Trial_1$ & \multicolumn{1}{l|}{\cellcolor[HTML]{FFF4D2}$y_1^1$} & \multicolumn{1}{l|}{\cellcolor[HTML]{FFF4D2}$y_2^1$} & \multicolumn{1}{l|}{\cellcolor[HTML]{FFF4D2}...} & \multicolumn{1}{l|}{\cellcolor[HTML]{FFF4D2}$y_i^1$} & \multicolumn{1}{l|}{\cellcolor[HTML]{FFF4D2}$y_{i+1}^1$} & \multicolumn{1}{l|}{\cellcolor[HTML]{FFF4D2}...} & $y_n^1$ \\ \hline 

\rowcolor[HTML]{FFF4D2} 
\cellcolor[HTML]{E5D1FA}$Trial_2$ & \multicolumn{1}{l|}{\cellcolor[HTML]{FFF4D2}$y_1^2$} & \multicolumn{1}{l|}{\cellcolor[HTML]{FFF4D2}$y_2^2$} & \multicolumn{1}{l|}{\cellcolor[HTML]{FFF4D2}...} & \multicolumn{1}{l|}{\cellcolor[HTML]{FFF4D2}$y_i^2$} & \multicolumn{1}{l|}{\cellcolor[HTML]{FFF4D2}$y_{i+1}^2$} & \multicolumn{1}{l|}{\cellcolor[HTML]{FFF4D2}...} & $y_n^2$ \\ \hline 

\rowcolor[HTML]{FFF4D2} 
\cellcolor[HTML]{E5D1FA}... & \multicolumn{1}{l|}{\cellcolor[HTML]{FFF4D2}...} & \multicolumn{1}{l|}{\cellcolor[HTML]{FFF4D2}...} & \multicolumn{1}{l|}{\cellcolor[HTML]{FFF4D2}...} & \multicolumn{1}{l|}{\cellcolor[HTML]{FFF4D2}...} & \multicolumn{1}{l|}{\cellcolor[HTML]{FFF4D2}...} & \multicolumn{1}{l|}{\cellcolor[HTML]{FFF4D2}...} & ... \\ \hline 

\rowcolor[HTML]{FFF4D2} 
\cellcolor[HTML]{E5D1FA}$Trial_j$ & \multicolumn{1}{l|}{\cellcolor[HTML]{FFF4D2}$y_1^j$} & \multicolumn{1}{l|}{\cellcolor[HTML]{FFF4D2}$y_2^j$} & \multicolumn{1}{l|}{\cellcolor[HTML]{FFF4D2}...} & \multicolumn{1}{l|}{\cellcolor[HTML]{FFF4D2}$y_i^j$} & \multicolumn{1}{l|}{\cellcolor[HTML]{FFF4D2}$y_{i+1}^j$} & \multicolumn{1}{l|}{\cellcolor[HTML]{FFF4D2}...} & $y_n^j$ \\ \hline 

\rowcolor[HTML]{FFF4D2} 
\cellcolor[HTML]{E5D1FA}$Trial_{j+1}$ & \multicolumn{1}{l|}{\cellcolor[HTML]{FFF4D2}$y_1^{j+1}$} & \multicolumn{1}{l|}{\cellcolor[HTML]{FFF4D2}$y_2^{j+1}$} & \multicolumn{1}{l|}{\cellcolor[HTML]{FFF4D2}...} & \multicolumn{1}{l|}{\cellcolor[HTML]{FFF4D2}$y_i^{j+1}$} & \multicolumn{1}{l|}{\cellcolor[HTML]{FFF4D2}$y_{i+1}^{j+1}$} & \multicolumn{1}{l|}{\cellcolor[HTML]{FFF4D2}...} & $y_n^{j+1}$ \\ \hline 

\rowcolor[HTML]{FFF4D2} 
\cellcolor[HTML]{E5D1FA}... & \multicolumn{1}{l|}{\cellcolor[HTML]{FFF4D2}...} & \multicolumn{1}{l|}{\cellcolor[HTML]{FFF4D2}...} & \multicolumn{1}{l|}{\cellcolor[HTML]{FFF4D2}...} & \multicolumn{1}{l|}{\cellcolor[HTML]{FFF4D2}...} & \multicolumn{1}{l|}{\cellcolor[HTML]{FFF4D2}...} & \multicolumn{1}{l|}{\cellcolor[HTML]{FFF4D2}...} & ... \\ \hline 

\rowcolor[HTML]{FFF4D2} 
\cellcolor[HTML]{E5D1FA}$Trial_{10}$ & \multicolumn{1}{l|}{\cellcolor[HTML]{FFF4D2}$y_1^{10}$} & \multicolumn{1}{l|}{\cellcolor[HTML]{FFF4D2}$y_2^{10}$} & \multicolumn{1}{l|}{\cellcolor[HTML]{FFF4D2}...} & \multicolumn{1}{l|}{\cellcolor[HTML]{FFF4D2}$y_i^{10}$} & \multicolumn{1}{l|}{\cellcolor[HTML]{FFF4D2}$y_{i+1}^{10}$} & \multicolumn{1}{l|}{\cellcolor[HTML]{FFF4D2}...} & $y_n^{10}$ \\ \hline 

& \multicolumn{7}{l|}{\cellcolor[HTML]{FFF4D2}Resampled objective function value ($lMSE$)} \\ 

\end{tabular}
\end{ruledtabular}
\end{table}

\nomenclature{\(\partial_n\)}{Partial derivative with respect to $n^{th}$ coordinate}
\nomenclature{\(p\)}{Pressure}
\nomenclature{\(U\)}{Mean velocity}
\nomenclature{\(x\)}{x-coordinate, input feature}
\nomenclature{\(U_e\)}{Equilibrium velocity}
\nomenclature{\(U^*\)}{Pseudo velocity}
\nomenclature{\(y\)}{y-coordinate, output label}
\nomenclature{\(\Omega_z\)}{Vorticity component in $z$ (spanwise) direction}
\nomenclature{\(\tau_w\)}{Wall shear stress}
\nomenclature{\(\rho\)}{Density at wall}
\nomenclature{\(\nu\)}{Dynamic viscosity}
\nomenclature{\(\delta\)}{Boundary layer thickness}
\nomenclature{\(\Pi\)}{Wake strength parameter}
\nomenclature{\(\delta^*\)}{Displacement thickness}
\nomenclature{\(\theta\)}{Momentum thickness}
\nomenclature{\(\tau\)}{Time}
\nomenclature{\(t\)}{Time (dummy variable)}
\nomenclature{\(\omega\)}{angular frequency $\omega = 2 \pi f$}
\nomenclature{\(c\)}{Speed of sound at free stream}
\nomenclature{\(\overline{p}\)}{Mean pressure}
\nomenclature{\(p^\prime\)}{Fluctuating pressure}
\nomenclature{\(\Phi_{pp}\)}{Power spectral density}
\nomenclature{\(R\)}{Auto-correlation function of pressure}
\nomenclature{\(\Delta\)}{Zagarola-Smits’s parameter}
\nomenclature{\(H\)}{Shape factor}
\nomenclature{\(M\)}{Mach Number}
\nomenclature{\(C_f\)}{Friction coefficient}
\nomenclature{\(R_T\)}{Outer-to-inner-layer timescale ratio}
\nomenclature{\(\beta\)}{Clauser parameter}
\nomenclature{\(m\)}{Slope}
\nomenclature{\(d\)}{Euclidean distance}
\nomenclature{\(i,j\)}{indices}
\nomenclature{\(n,N\)}{Length of a vector}
\nomenclature{\(dB\)}{Decibel}
\nomenclature{\(w\)}{Weights for the neuron inputs}
\nomenclature{\(b\)}{bias for neuron}
\nomenclature{\(lMSE\)}{Logarithmic mean squared error}
\nomenclature{\(W\)}{Weights for labels in objective funcion}
\nomenclature{\(h\)}{Head length}
\nomenclature{\(P_{opt}\)}{Optimization probability}
\nomenclature{\(f\)}{Frequency}
\nomenclature{\(A_{log}\)}{Logarithmic amplitude}
\nomenclature{\(A_{lin}\)}{Linear amplitude}
\nomenclature{\(MSE\)}{Mean squared error}
\nomenclature{\(O(n)\)}{Order of n operations}
\nomenclature{\(gen\)}{Number of generations}
\nomenclature{\(Y^{GEP}\)}{Output of GEP linking function}
\nomenclature{\(Y^{true}\)}{Labels used to train a ML algorithm}
\nomenclature{\(<y>_E\)}{Ensemble average of the batch of outputs}
\nomenclature{\(median(y)\)}{Median of the batch of outputs}
\nomenclature{\(y_{25-75}\)}{Outputs in the interquartile range of the batch of outputs}
\nomenclature{\(y_{max}\)}{Maximum value among the batch of outputs}
\nomenclature{\(y_{min}\)}{Minimum value among the batch of outputs}

\printnomenclature

\section*{References}

\nocite{*}
\bibliography{references}

\end{document}